\documentclass[pre,superscriptaddress,twocolumn,showkeys]{revtex4-1}
\newcommand{\h}{\mathcal{H}(\lambda)}
\newcommand{\hh}{\mathcal{H}}
\newcommand{\heff}{\mathcal{H}_{\footnotesize{\textrm{eff}}}}

\newcommand{\hmin}{\mathcal{H}_{\min}}
\newcommand{\hmax}{\mathcal{H}_{\max}}
\newcommand{\seff}{S_{\footnotesize{\textrm{eff}}}}

\newcommand{\ceff}{\eta_{\footnotesize{\textrm{eff}}}}

\newcommand{\lmin}{\ell^{A}(\Lambda_{\min})}
\newcommand{\lmax}{\ell^{A}(\Lambda_{\max})}
\newcommand{\heffnl}{\mathcal{H}_{\footnotesize{\textrm{eff}}}}

\usepackage{amsmath,amssymb,graphicx,subfigure,multirow,hyperref,url}
\usepackage{color}

\begin{document}
\title[Taxonomies of networks]{Taxonomies of networks from community structure}

\author{Jukka-Pekka Onnela}\thanks{These authors contributed equally to this work.}
\affiliation{Department of Biostatistics, Harvard School of Public Health, Boston, MA 02115, USA}
\affiliation{Department of Health Care Policy, Harvard Medical School, Boston, MA 02115, USA}
\affiliation{Department of Physics, University of Oxford, Oxford OX1 3PU, UK}
\affiliation{CABDyN Complexity Centre, University of Oxford, Oxford OX1 1HP, UK}

\author{Daniel J. Fenn}\thanks{These authors contributed equally to this work.}
\affiliation{Mathematical and Computational Finance Group, University of Oxford, Oxford OX1 3LB, UK}
\affiliation{CABDyN Complexity Centre, University of Oxford, Oxford OX1 1HP, UK}

\author{Stephen Reid}
\affiliation{Department of Physics, University of Oxford, Oxford OX1 3PU, UK}

\author{Mason A. Porter}
\affiliation{Oxford Centre for Industrial and Applied Mathematics, Mathematical Institute, University of Oxford, OX1 3LB, UK}
\affiliation{CABDyN Complexity Centre, University of Oxford, Oxford OX1 1HP, UK}

\author{Peter J. Mucha}
\affiliation{Carolina Center for Interdisciplinary Applied Mathematics, Department of Mathematics and Institute for Advanced Materials, Nanoscience \& Technology, University of North Carolina, Chapel Hill, NC 27599, USA}

\author{Mark D. Fricker}
\affiliation{Department of Plant Sciences, University of Oxford, South Parks Road, Oxford, OX1 3RB, UK}
\affiliation{CABDyN Complexity Centre, University of Oxford, Oxford OX1 1HP, UK}

\author{Nick S. Jones}
\affiliation{Department of Physics, University of Oxford, Oxford OX1 3PU, UK}
\affiliation{Oxford Centre for Integrative Systems Biology, Department of Biochemistry, University of Oxford, Oxford, OX1 3QU, UK}
\affiliation{CABDyN Complexity Centre, University of Oxford, Oxford OX1 1HP, UK}

\begin{abstract}
The study of networks has become a substantial interdisciplinary endeavor that encompasses myriad disciplines in the natural, social, and information sciences. Here we introduce a framework for constructing taxonomies of networks based on their structural similarities. These networks can arise from any of numerous sources: they can be empirical or synthetic, they can arise from multiple realizations of a single process (either empirical or synthetic), they can represent entirely different systems in different disciplines, etc. Because mesoscopic properties of networks are hypothesized to be important for network function, we base our comparisons on summaries of network community structures. Although we use a specific method for uncovering network communities, much of the introduced framework is independent of that choice. After introducing the framework, we apply it to construct a taxonomy for 746 networks and demonstrate that our approach usefully identifies similar networks.  We also construct taxonomies within individual categories of networks, and we thereby expose nontrivial structure. For example, we create taxonomies for similarity networks constructed from both political voting data and financial data.  We also construct network taxonomies to compare the social structures of 100 Facebook networks and the growth structures produced by different types of fungi.
\end{abstract}


\maketitle

\section{Introduction}

Although there is a long tradition of scholarship on networks, the last two decades have witnessed substantial advances in network science due to developments in physics, mathematics, computer science, sociology, and numerous other disciplines \cite{newman2010,newmanphystoday}. Given that the questions asked by researchers in different fields can be surprisingly similar, it would be useful to be able to highlight similarities in network structures across disciplines in a systematic way. One way to approach this is to formulate a suitable means of comparing networks and to use this means to develop taxonomies of networks. Such taxonomies have the potential to facilitate the identification of problems from different disciplines that might be approached similarly in terms of both empirical analyses and theoretical modeling. For example, if a biological network depicting covariation of neural activity in different regions of the brain is demonstrated to be structurally similar to a financial network representing correlations of stock returns, then certain types of edge thresholding methods or structural null models might be applicable to both situations.

From a historical perspective, classification of objects has often been central to the progress of science, as demonstrated by the periodic table of elements in chemistry and phylogenetic trees of organisms in biology \cite{borner2010}.  It is plausible that an organization of networks has the potential to shed light on mechanisms for generating networks, reveal how an unknown network should be treated once one has discerned its position in a taxonomy, or help identify a network family's anomalous members. Further potential applications of network taxonomies include unsupervised study of multiple realizations of a given model process (e.g., characterizing the similarities and differences of many different networks drawn from the Erd\"os-R\'enyi random graph model using the same parameter values), examination of multiple empirical networks with known similar origins or generative processes, and the detection of anomalous changes in temporally ordered series of networks. In this paper, we develop a framework for the creation of network taxonomies \footnote{The code used for constructing network taxonomies in this paper is available at {\tt www.jponnela.com}.}. In so doing, we develop the requisite diagnostic tools and discuss several case studies that suggest how our methodology can help illuminate relationships both between and within families of networks.

In aiming to construct taxonomies of networks, one has to consider the scales at which one wants to compare differences in network structures. Much research has focused on extremes---either microscopic (e.g., node degree) or macroscopic (e.g., mean geodesic distance) properties---and numerous researchers have, for example, reported that many empirical networks possess heavy-tailed degree distributions or the small-world property \cite{barabasirev,newman2010}. Given the ubiquity of such findings, it is clear that more nuanced approaches are needed to make useful comparisons between networks. Indeed, interpretations of microscopic and macroscopic approaches often implicitly assume that networks are homogeneous and ignore ``mesoscopic'' structures in networks. To overcome some of these limitations, earlier work has focused on the statistics of small, \emph{a priori} specified modules called ``motifs'' \cite{motifs,superfamily}, role-to-role connectivity profiles of nodes \cite{profiles}, the isolation of statistically significant structures called ``backbones'' \cite{backbone}, interrelations of network modules \cite{optimalmap},  examination of the number of nodes located within ``shells'' \cite{bagrow2008}, and the self-similarity of networks as characterized by fractal exponents \cite{song}. The taxonomic framework that we develop in the present paper builds on the idea of examining network modules by computing community structures \cite{comnotices,fortunato2009}, as was also done in the work of \cite{fortunato_comprop}, and we subsequently compare signatures derived from community structure across networks. Importantly, although we use a specific method to uncover network communities, much of the introduced framework is independent of that choice. Consequently, our comparative framework can accommodate a large variety of community detection schemes.

The remainder of this paper is organized as follows.  First, we discuss the detection of communities in networks in order to find coherent groups of nodes that are densely connected to each other.  We then introduce \emph{mesoscopic response functions} (MRFs), which allow us to probe how the community structure of a network changes as a function of a resolution parameter that determines network scales of interest.  We then illustrate MRFs using several examples of networks and compare the MRFs for several well-known generative models of networks.  We use MRFs to develop a means to measure distance between a pair of networks, and use this comparative measure to cluster networks and thereby develop taxonomies.  Using 746 networks from numerous different fields, we construct a taxonomy of these networks. We then construct taxonomies of networks within fields using several case studies: voting in the United States Senate, voting in the United Nations General Assembly, Facebook networks at US universities, fungal networks, and networks of stock returns in the New York Stock Exchange. In each example, we expose structure that is either illuminating or can be checked against information from an external source (e.g., previously published investigations). This suggests that our method for comparing networks is capturing important similarities and differences.  We conclude with a brief summary and discussion of our results. In addition, we provide further details in the Appendices and Supplemental Material.  Among other topics, we examine the robustness of the obtained taxonomies, address some computational issues, tabulate some of the basic properties of the networks that we investigated, and provide references for the network data sources used in this study.

\section{Multi-Resolution Community Detection}

Our approach is based on network \emph{community structure} \cite{comnotices,fortunato2009}. A community consists of a set of nodes for which there are more edges (or, in the case of weighted networks, a greater total edge weight) connecting the nodes in the set than what would be expected by chance.  The algorithmic detection of communities is a particularly active area of network science, in part because communities are thought to be related to functional units in many networks and in part because they can strongly influence dynamical processes that operate on networks \cite{comnotices,fortunato2009}.

In this paper, we detect communities using the multi-resolution Potts method \cite{potts2,comnotices,fortunato2009}, a generalization of modularity optimization \cite{newmangirvan,newmanpnas2006,potts1,potts2,comnotices,fortunato2009}.  (Modularity optimization is perhaps the most popular approach for detecting communities.)  Given a network adjacency matrix $A_{ij}$, we find communities by minimizing the Hamiltonian of the infinite-range $N$-state Potts spin glass
\begin{align}
	\h &= - \sum_{i \ne j} J_{ij}(\lambda) \delta(C_i,C_j) \notag \\
	 &= - \sum_{i \ne j} \left( A_{ij} - \lambda P_{ij} \right)\delta(C_i,C_j)\,, \label{eq:SOM:H}
\end{align}
where $C_i$ indicates the community (state) of node (spin) $i$, $\lambda$ is a resolution parameter, and $\mathbf{J}(\lambda)$ is the coupling matrix with entries $J_{ij}(\lambda)$ representing the interaction strength between node $i$ and node $j$ in the Potts Hamiltonian.  We use the (undirected-network) null model $P_{ij} = k_ik_j/(2m)$, where $k_i$ denotes the strength (total edge weight) of node $i$ and $m$ is the total edge weight in the network \cite{newmangirvan}. By tuning the resolution parameter $\lambda$, we can detect communities at multiple scales of a network.  Our particular choice of $J_{ij}$ implies that we are optimizing modularity (with the addition of the resolution parameter) \cite{comnotices,fortunato2009}.

To compare networks, we create profiles of summary statistics that characterize the community structure of each network at different mesoscopic scales. We also study a wide variety of networks that contain different numbers of nodes and edges.  (We enumerate the networks that we consider in Table~II of the Supplemental Material.) To ensure that we can compare the profiles for different networks, we sweep the resolution parameter $\lambda$ from a minimum value $\Lambda_{\min}$ to a maximum value $\Lambda_{\max}$ (discussed in detail below). We define these quantities separately for each network such that the number of communities $\eta$ into which the network is partitioned is $1$ at $\Lambda_{\min}$ and is equal to the total number of nodes $N$ at $\Lambda_{\max}$. In other words, one can think of $\lambda$ as a parameter that controls the fragmentation of a network into communities.

To find the minimum and maximum resolution-parameter values, consider the interactions in Eq.~(\ref{eq:SOM:H}).  An interaction is called \emph{ferromagnetic} when $J_{ij}>0$ and \emph{antiferromagnetic} when $J_{ij}<0$.  For each pair of nodes $i$ and $j$, we find the resolution $\lambda=\Lambda_{ij}$ at which the interaction $J_{ij}$ is neutral (i.e., $J_{ij}(\Lambda_{ij}) = 0$), leading to $\Lambda_{ij} = A_{ij} / P_{ij}$. We thereby identify two special resolutions:
\begin{align}
	\Lambda_{\min} &= \max_{ij} \left\{ \Lambda_{ij} | \eta(\lambda) =1 \right\}\,,\\
	\Lambda_{\max} &= \max_{ij} \left\{\Lambda_{ij}\right\}+\epsilon\,,
\end{align}
where $\epsilon>0$ is any small number (we use $\epsilon = 10^{-6}$ in the present paper).  The resolution $\Lambda_{\min}$ is the largest $\Lambda_{ij}$ value for which community detection yields a single community; note that this need not be the minimum non-zero value of $\Lambda_{ij}$. Including the small number $\epsilon$ in the definition of $\Lambda_{\max}$ ensures that all edges are antiferromagnetic at resolution $\lambda=\Lambda_{\max}$ and thereby forces each node into its own community.

\section{Mesoscopic Response Functions (MRFs)}

\begin{figure}[htp]
\begin{center}
\subfigure{\includegraphics[width=1\linewidth]{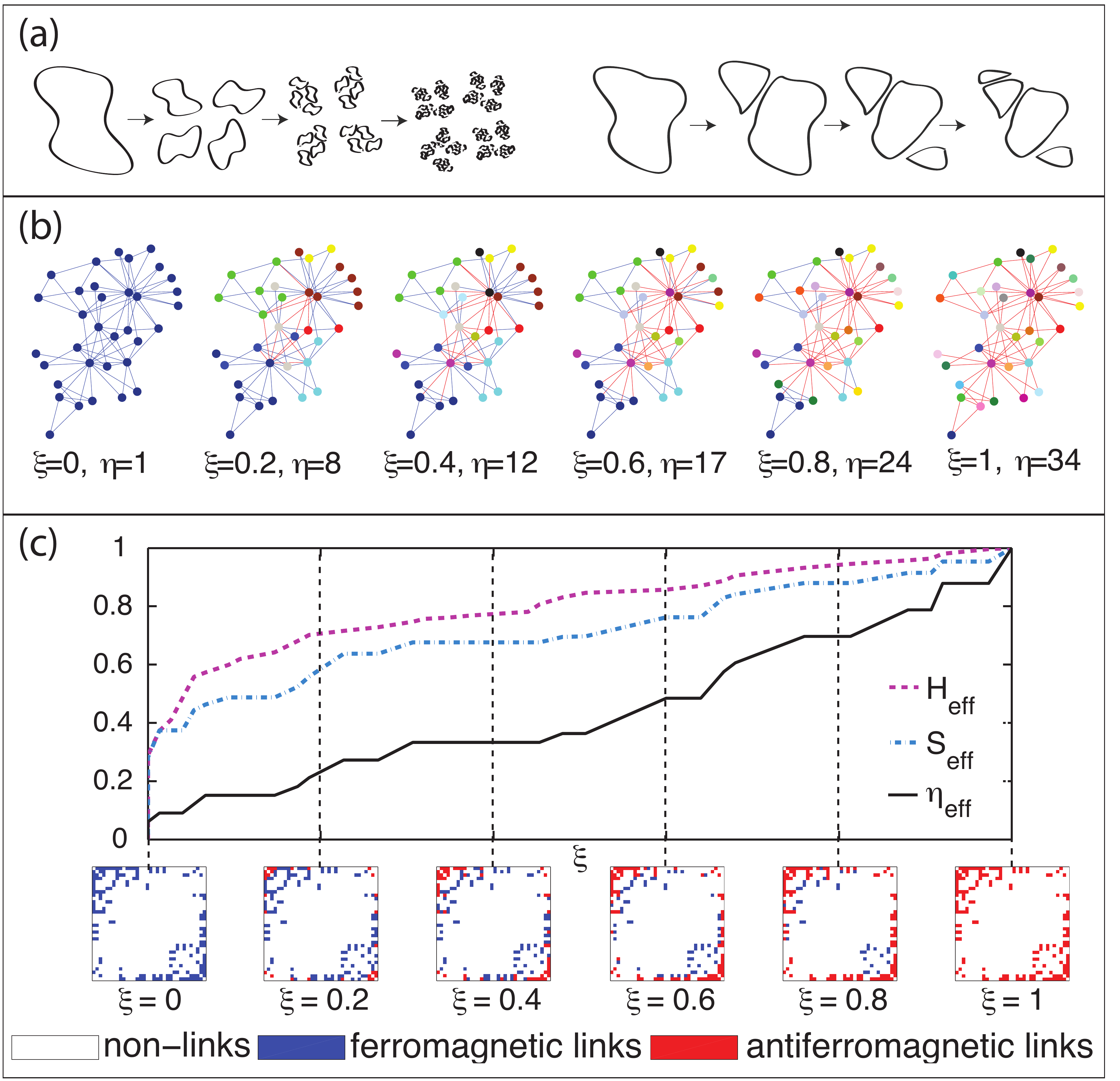}}
\caption{(Color online)
(\textbf{a}) Schematic of some of the ways that a network can break up into communities as the value of $\lambda$ (or $\xi$) is increased.
(\textbf{b}) Zachary Karate Club network \cite{karate} for different values of the effective fraction of antiferromagnetic edges $\xi$. All interactions are either ferromagnetic or antiferromagnetic, i.e. for the values of $\xi$ used, there are no neutral interactions. We color edges in blue if the corresponding interactions are ferromagnetic, and we color them red if the interactions are antiferromagnetic. We color the nodes based on community affiliation.
(\textbf{c}) The $\heff$, $\seff$, and $\ceff$ MRFs, and the interaction matrix $\mathbf{J}$ for different values of $\xi$. We color elements of the interaction matrix by depicting the absence of an edge in white, ferromagnetic edges in blue, and antiferromagnetic edges in red.}
\label{karate_MRFs}
\end{center}
\end{figure}

To describe how a network disintegrates into communities as the value of $\lambda$ is increased from $\Lambda_{\min}$ to $\Lambda_{\max}$ (see Fig.~\ref{karate_MRFs}(a) for a schematic), one needs to select summary statistics. There are many possible ways to summarize such a disintegration process, and we focus on three diagnostics that characterize fundamental properties of network communities.

First, we use the value of the Hamiltonian $\hh(\lambda)$ (\ref{eq:SOM:H}), which is a scalar quantity closely related to network modularity and quantifies the energy of the system \cite{comnotices,fortunato2009}. Second, we calculate a partition entropy $S(\lambda)$ to characterize the community size distribution.  To do this, let $n_k$ denote the number of nodes in community $k$ and define $p_{k} = n_{k}/N$ to be the probability to choose uniformly at random a member node of community $k$.  This yields a (Shannon) partition entropy of $S(\lambda) = -\sum_{k=1}^{\eta(\lambda)} p_{k} \log{p_{k}}$, which quantifies the disorder in the associated community size distribution. Third, we use the number of communities $\eta(\lambda)$.

Needing to normalize $\hh$, $S$, and $\eta$ to compare them effectively across networks, we define an \emph{effective energy}
\begin{equation}
	\heff(\lambda) = \frac{\h-\hmin}{\hmax-\hmin} = 1 - \frac{\h}{\hmin}\,,
\end{equation}
where $\hmin = \hh(\Lambda_{\min})$ and $\hmax = \hh(\Lambda_{\max})$; 
an \emph{effective entropy}
\begin{equation}
	\seff(\lambda) = \frac{S(\lambda)-S_{\min}}{S_{\max}-S_{\min}} = \frac{S(\lambda)}{\log{N}}\,,
\label{eg:seffeq}
\end{equation}
where $S_{\min} = S(\Lambda_{\min})$ and $S_{\max} = S(\Lambda_{\max})$;
and an \emph{effective number of communities}
\begin{equation}
	\ceff(\lambda) = \frac{\eta(\lambda) - \eta_{\min}}{\eta_{\max} - \eta_{\min}} = \frac{\eta(\lambda) - 1}{N-1}\,,
\end{equation}
where $\eta_{\min} = \eta(\Lambda_{\min})$ and $\eta_{\max} = \eta(\Lambda_{\max})$.

Some networks contain a small number of entries $\Lambda_{ij}$ that are orders-of-magnitude larger than most other entries. For example, in the network of Facebook friendships at Caltech \cite{traud,datadump}, 98\% of the $\Lambda_{ij}$ entries are less than 100, but 0.02\% of them are larger than 8000. These large $\Lambda_{ij}$ values arise when two low-strength nodes become connected.  Using the null model $P_{ij} = k_ik_j/(2m)$, the interaction between two nodes $i$ and $j$ becomes antiferromagnetic when $\lambda>A_{ij}/P_{ij}=2mA_{ij}/(k_ik_j)$. If the network has a large total edge weight but both $i$ and $j$ have small strengths compared to other nodes in the network, then $\lambda$ needs to be large to make the interaction antiferromagnetic. In prior studies, network community structure has been investigated at different mesoscopic scale by considering plots of various diagnostics as a function of the resolution parameter \cite{comnotices,fortunato2009,potts2}.  In the present example, such plots would be dominated by interactions that require large resolution-parameter values to become antiferromagnetic. To overcome this issue, we define the \textit{effective fraction of antiferromagnetic edges}
\begin{equation}
	\xi = \xi(\lambda) = \frac{\ell^A (\lambda) - \lmin}{\lmax - \lmin} \in [0,1] \,, \label{eq:SOM:xi}
\end{equation}
where $\ell^A (\lambda)$ is the total number of antiferromagnetic interactions for the given value of $\lambda$ in the network. In other words, it is the number of $\Lambda_{ij}$ elements that are smaller than $\lambda$. Thus, $\lmin$ is the largest number of antiferromagnetic interactions for which the network still forms a single community, and the effective number of antiferromagnetic interactions $\xi(\lambda)$ is the number of antiferromagnetic interactions (normalized to the unit interval) in excess of $\lmin$. The function $\xi(\lambda)$ increases monotonically in $\lambda$.

\begin{figure}[htp]
\begin{center}
\includegraphics[width = 0.8\linewidth]{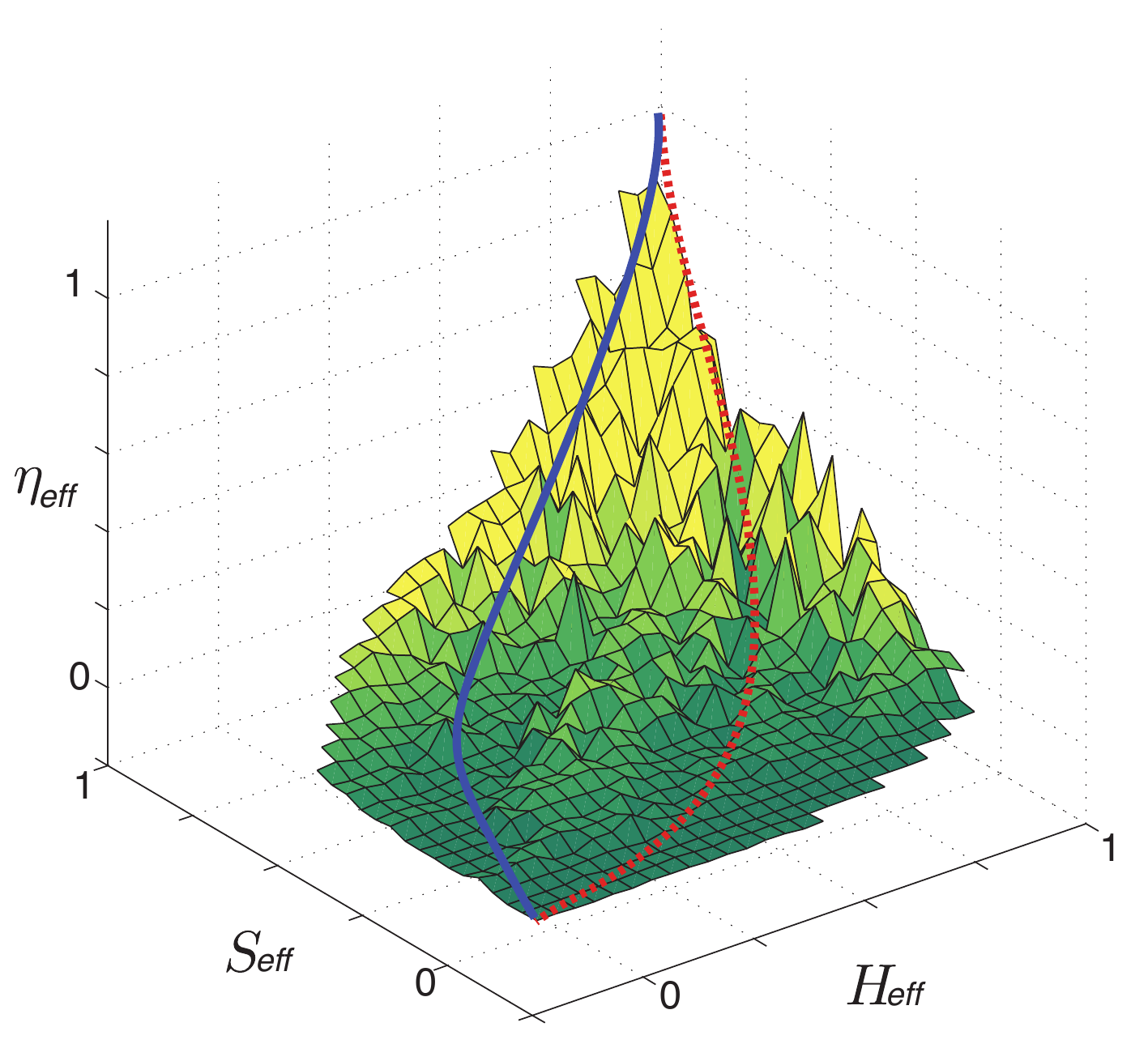}
\caption{(Color online)
The mesoscopic response function (MRF) of a given network consists of a 3-dimensional vector ($\heff(\xi)$, $\seff(\xi)$, $\ceff(\xi)$), where $\xi\in[0,1]$. By construction, the MRF starts from the bottom front corner [$\heff(\xi=0)$, $\seff(\xi=0)$, $\ceff(\xi=0)$] and ends at the top back corner [$\heff(\xi=1)$, $\seff(\xi=1)$, $\ceff(\xi=1)$]. The colored surface plot shows where most MRFs lie. We also show schematic MRFs in blue (solid curve) and red (dashed curve).}
\label{fig:cube}
\end{center}
\end{figure}

Sweeping $\lambda$ from $\Lambda_{\min}$ to $\Lambda_{\max}$ corresponds to sweeping the value of $\xi$ from $0$ to $1$.  (One can think of $\lambda$ as a continuous variable and $\xi$ as a discrete variable that changes with events.)  As we perform such sweeping for a given network, the number of communities increases from $\eta(\xi = 0) = 1$ to $\eta(\xi = 1) = N$ and yields a vector ($\heff(\xi)$, $\seff(\xi)$, $\ceff(\xi)$) whose components we call the \emph{mesoscopic response functions} (MRF) of that network. Because $\heff\in[0,1]$, $\seff\in[0,1]$, $\ceff\in[0,1]$, and $\xi\in[0,1]$ for every network, we can compare the MRFs across networks and use them to identify groups of networks with similar mesoscopic structures. In Fig.~\ref{karate_MRFs}(b), we show the Zachary Karate Club network \cite{karate} for different values of $\xi$. As more edges become antiferromagnetic, the network fragments into smaller communities, and panel (c) shows the corresponding MRFs. In Fig.~\ref{fig:cube}, we show a schematic of the MRF in which we emphasize its interpretation as a 3-dimensional vector. In Fig.~\ref{fig:MRF_eg}, we show example MRFs for several other networks.

Although minimizing Eq.~(\ref{eq:SOM:H}) is an NP-hard problem \cite{BRANDES_2006} and $\mathcal{H}$ possesses a complicated landscape of local optima for many networks \cite{modcaution}, there exist numerous good computational heuristics that make finding a nearly-optimal partition of the network into communities at a given resolution computationally tractable \cite{comnotices,fortunato2009}. Thus far, we have reported results that were obtained by optimizing modularity using the locally greedy Louvain algorithm \cite{blondel} because its speed was important for studying large networks. We have compared the results that we report in the present work to those obtained from optimizing modularity using spectral and simulated-annealing algorithms, and obtained similar MRFs and taxonomies for them (see Appendix~\ref{sec:heuristics}2 for more details).

\section{Examples of MRFs}

The shapes of the MRFs summarize many factors---including the fraction of possible edges in a network that are actually present, the relative weights of inter- versus intra-community edges, the edge weights compared with the expected edge weights in the null model, the number of edges that need to become antiferromagnetic for a community to fragment, and the way in which the communities fragment (e.g., whether a community splits in half or a single node leaves a community when a particular edge becomes antiferromagnetic). To understand the effects of some of these factors on the shapes of the MRFs, we consider some examples.

\begin{figure}[htp]
\begin{center}
\includegraphics[width=1\linewidth]{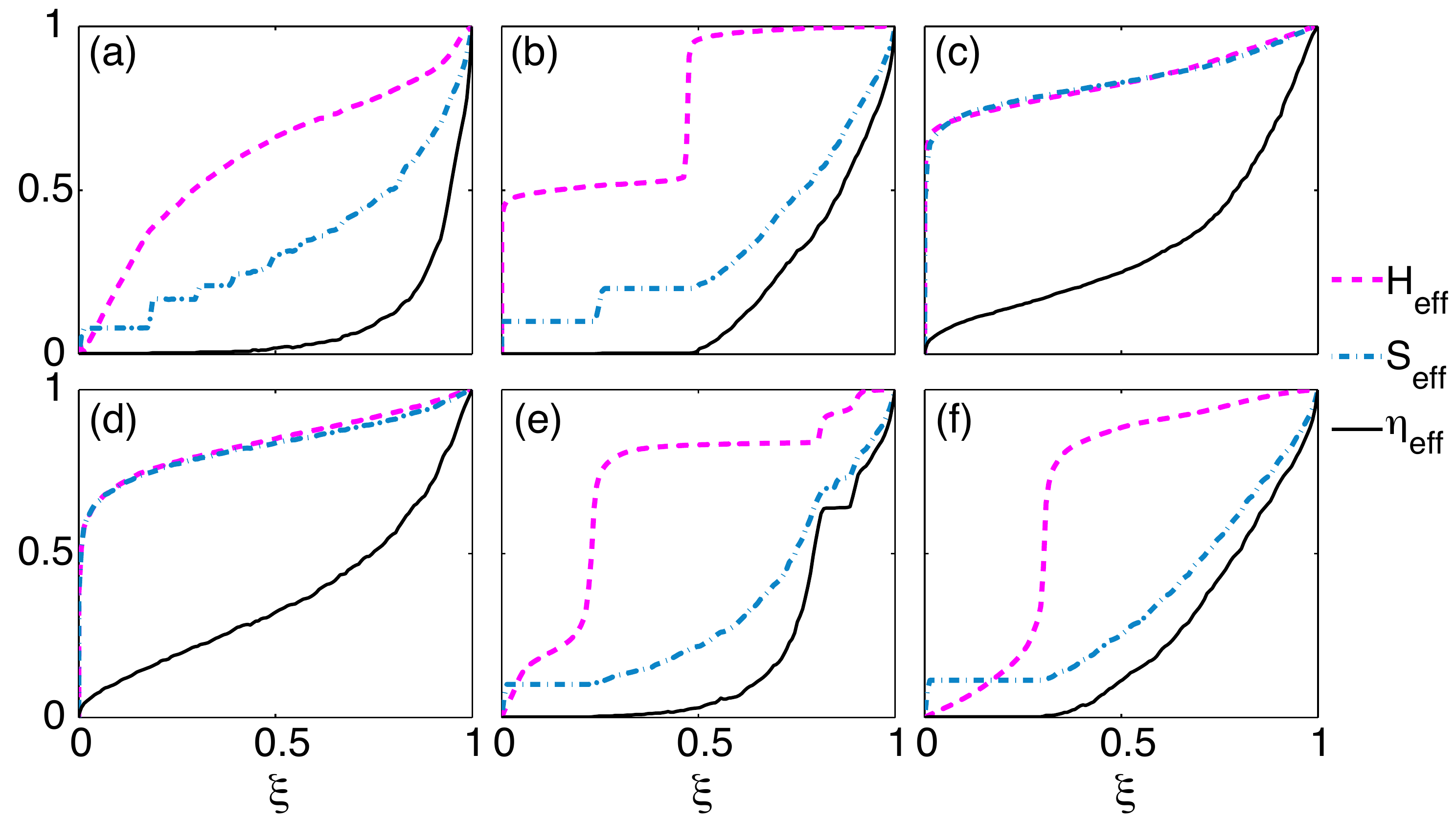}
\caption{(Color online) Example mesoscopic response functions (MRFs). The curves show $\heffnl$ (pink, dashed), $\seff$ (blue, dash-dotted), and $\ceff$ (black, solid) as a function of the effective fraction of antiferromagnetic edges $\xi$ for the following networks:
(\textbf{a}) New York Stock Exchange (NYSE), 1980--1999 \cite{nyse};
(\textbf{b}) Fractal (10,2,8) \cite{fractal};
(\textbf{c}) Biogrid \emph{D. melanogaster} \cite{biogrid};
(\textbf{d}) Garfield scientometrics citations \cite{garfield};
(\textbf{e}) United Kingdom House of Commons voting, 2001--2005 \cite{commonsdata};
(\textbf{f}) Roll-call voting of 108th United States House of Representatives \cite{voteview,poole,mccarty,waugh}.}
\label{fig:MRF_eg}
\end{center}
\end{figure}

Of particular interest are plateaus in the $\ceff$ and $\seff$ curves that are accompanied by large increases in $\heff$.  As illustrated in panel \ref{fig:MRF_eg}(a), the New York Stock Exchange (NYSE) network from 1980 to 1999 \cite{nyse} provides a good example of this behavior. This network is an instance from the category of \emph{similarity networks}.  We use this label to describe networks that have been constructed by starting from some node-level quantity or attribute and then defining the edges based on some form of similarity or correlation measure between each pair of nodes. Similarity networks tend to be complete (or almost complete) and weighted networks, except when they have been deliberately thresholded. In this particular example, each node represents a stock, and the strength of the edge connecting stocks $i$ and $j$ is linear in the Pearson correlation between the daily logarithmic returns of the stocks. (See Section~\ref{subsec:nyse} for more details.) Plateaus imply that as the resolution $\lambda$ is increased (leading to an increase in $\heff$), the communities remain unchanged even though the number and strength of antiferromagnetic interactions increase. As $\lambda$ is increased and more interactions become antiferromagnetic, there is an increased energy incentive for communities to break up.  Community partitions in such plateaus tend to be robust and have the potential to represent interesting structures \cite{arenasrobust,potts2,comnotices,fortunato2009}.

In Fig.~\ref{fig:MRF_eg}(b), we show MRFs for a ``fractal" network \cite{fractal}, which demonstrates that plateaus in the $\ceff$ and $\seff$ curves need not be accompanied by significant changes in $\heffnl$. Such plateaus can be explained by considering the distribution of $\Lambda_{ij}$ values. If several interactions have identical values of $\Lambda_{ij}$, then the interactions all become antiferromagnetic at exactly the same resolution value. This leads to a significant increase in the effective fraction of antiferromagnetic edges $\xi$ but only a small change in $\heffnl$. If these interactions do not result in additional communities, then we obtain plateaus in the $\ceff$ and $\seff$ curves.

To demonstrate qualitatively different behavior, we show the MRFs for the Biogrid \emph{Drosophila melanogaster} network and the Garfield Scientometrics citation network in Fig.~\ref{fig:MRF_eg}(c) and Fig.~\ref{fig:MRF_eg}(d), respectively. A common feature in these MRFs is the sharp initial increase in the curves that results from the networks initially breaking into two communities.

Another family of networks, which we will discuss in more detail in our case studies, are political voting networks. These voting networks are also similarity networks: we have constructed these networks so that an edge between two nodes indicates the level of agreement on votes between two entities, and each edge takes a value between $0$ and $1$.  In Fig.~\ref{fig:MRF_eg}(e), we show the MRFs for the voting network of the United Kingdom House of Commons during the period 2001--2005 \cite{commonsdata}; in Fig.~\ref{fig:MRF_eg}(f), we show the MRFs for the roll-call voting network for the $108^{\textrm{th}}$ (2003--2004) United States House of Representatives \cite{voteview,poole,mccarty,waugh}.  In both cases, we observe that sharp increases in $\heffnl$ can be accompanied by only small changes in $\ceff$ and $\seff$. To see how this can arise, we again consider the distribution of $\Lambda_{ij}$ values. If the $\Lambda_{ij}$ distribution is multi-modal, there can be a large difference between consecutive $\Lambda_{ij}$ values. A large increase in $\lambda$ is then needed to increase $\xi$, which in turn results in a large change in $\heffnl$. However, the change in $\ceff$ is small because this only results in a single additional antiferromagnetic interaction.

\section{Comparing Network Models}

\begin{figure}[htp]
\begin{center}
\includegraphics[width=1\linewidth]{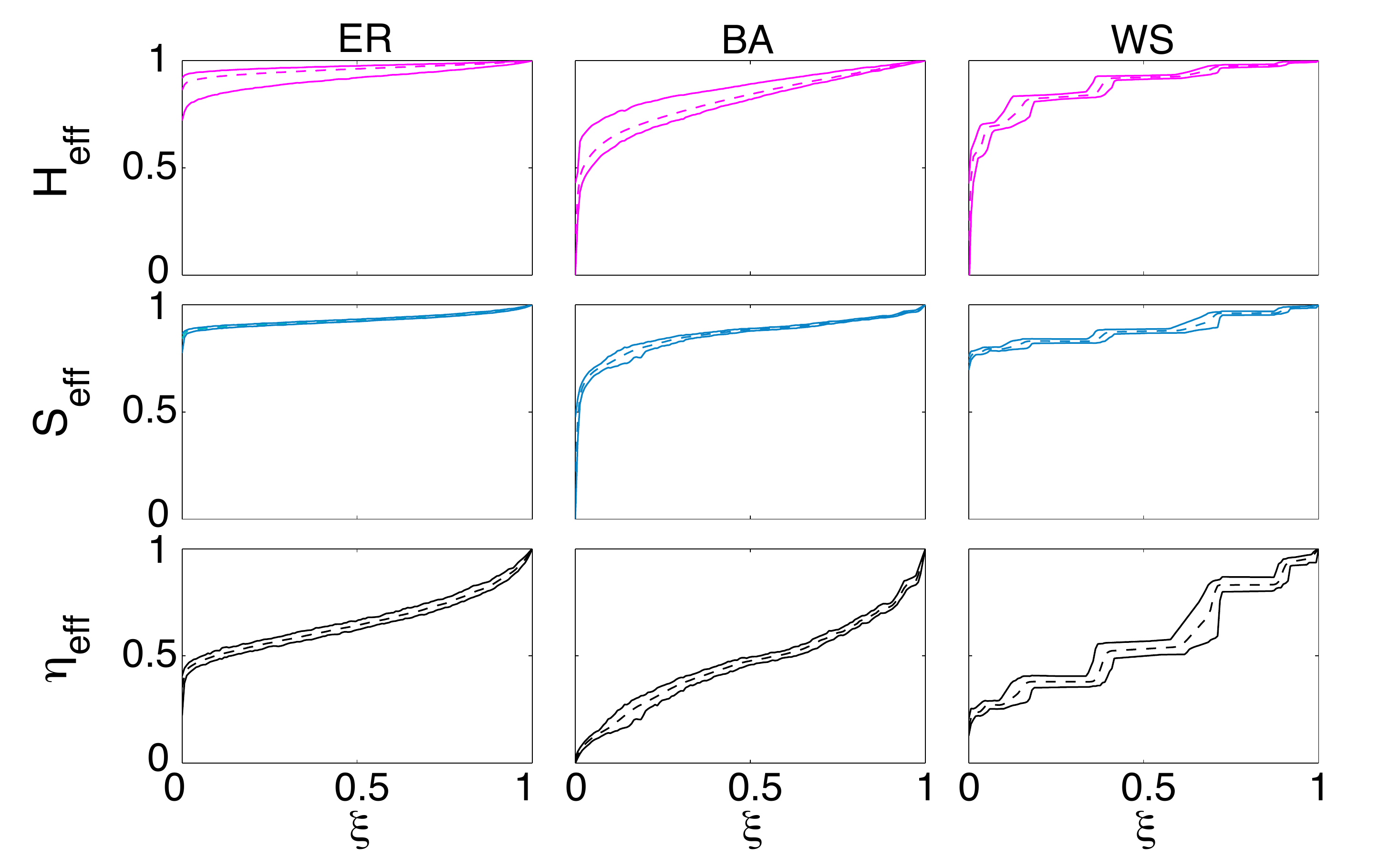}
\caption{(Color online) MRFs for 1000 realizations of Erd\"os-R\'enyi (ER), Barab\'asi-Albert (BA), and Watts-Strogatz (WS) networks. Each network has $N=1000$ nodes and mean degree $\langle k \rangle=10$. For each value of $\xi$, the upper curves show the maximum values of $\heffnl$ (top row), $\seff$ (middle row), and $\ceff$ (bottom row) for all networks in the ensemble; the lower curves show the corresponding minimum value, and the dashed curves show the corresponding mean.}
\label{fig:model_MRFs}
\end{center}
\end{figure}

To provide further insights into MRFs, we consider Erd\"os-R\'enyi (ER) \cite{er}, Barab\'asi-Albert (BA) \cite{ba}, and Watts-Strogatz (WS) \cite{ws} networks.  These network models are stochastic, and there is a large ensemble of possible network realizations for each choice of parameter values in these models. However, even with the ensuing structural variation, networks generated by a given one of these three models exhibit similar properties at mesoscopic and macroscopic scales, so we expect MRFs for different realizations of a given model to be similar. In Fig.~\ref{fig:model_MRFs}, we compare the MRFs for 1000 realizations of each model for networks with $N=1000$ nodes and mean degree $\langle k \rangle=10$.  For the WS networks, we set the edge rewiring probability at $p=0.1$.  As illustrated in Fig.~\ref{fig:model_MRFs}, we obtain a narrow range of possible MRFs for fixed parameter values.  This comparison illustrates that the MRF profiles of the three different models are distinctive. In addition, for each model there is little variation in the behavior of the MRFs across different network realizations with the same parameter values.

\begin{figure}[htp]
\begin{center}
\includegraphics[width=1\linewidth]{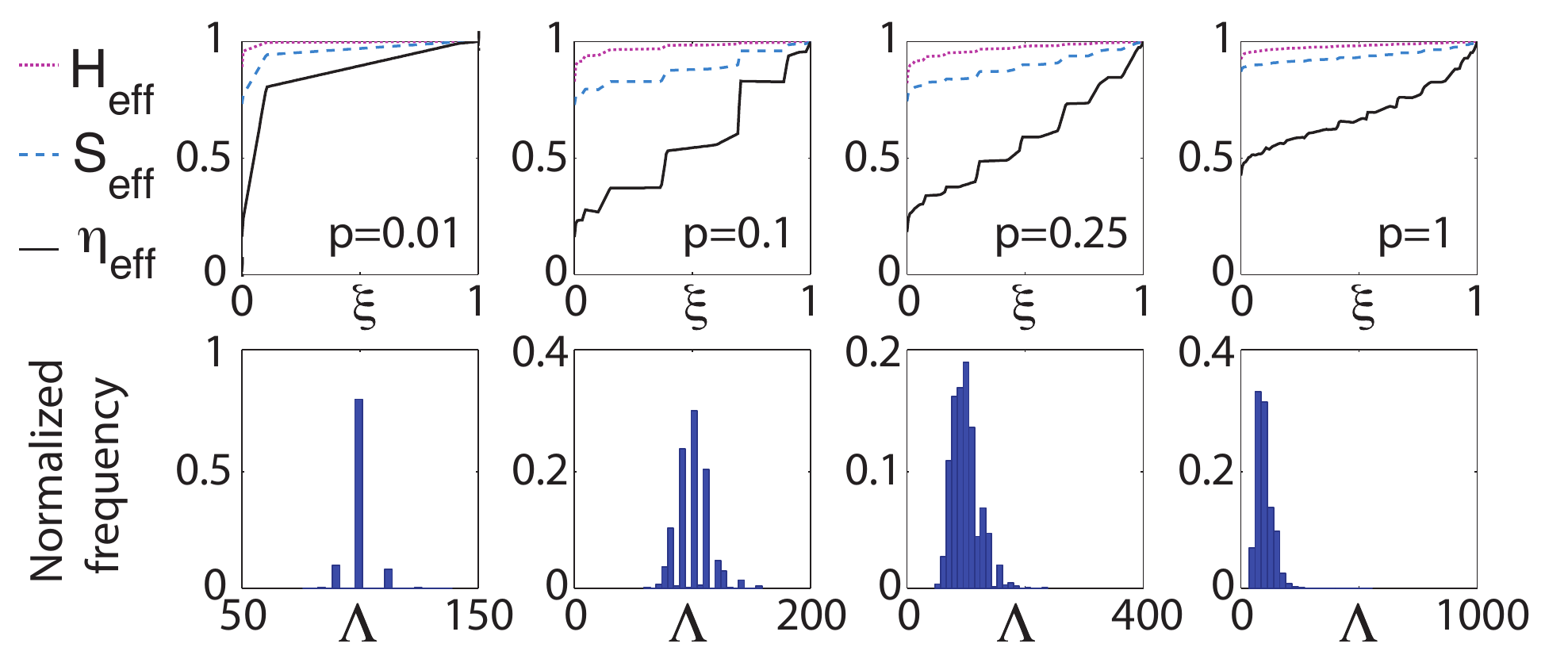}
\caption{(Color online) Upper panels: MRFs for Watts-Strogatz networks for different values of the rewiring probability $p$. Each network has $N=1000$ nodes and mean degree $\langle k \rangle=10$. Lower panels: distributions of $\Lambda_{ij}$ values for each network. As expected, the MRFs for $p=1$ are identical to those of an Erd\"os-R\'enyi network with $N=1000$ and $\langle k \rangle=10$.}
\label{fig:WS_comp}
\end{center}
\end{figure}

It is also instructive to consider variation in MRF shapes for a particular network model for different parameter values. We focus on WS networks because they illuminate the effect of the distribution of $\Lambda_{ij}$ values on the shapes of the MRFs. In Fig.~\ref{fig:WS_comp}, we show MRFs for WS networks for different values of the edge rewiring probability $p$. (We continue using $N=1000$ and $\langle k \rangle=10$.)  We also show the distribution of $\Lambda_{ij}$ values for each network.

For small rewiring probabilities, the MRFs have lots of steps. As with prior examples, we can see how this feature arises by considering the distribution of $\Lambda_{ij}$ values. When the rewiring probability is small, many nodes possess the same degree, which results in the presence of many interactions with identical $\Lambda_{ij}$ values (see the bottom left panel of Fig.~\ref{fig:WS_comp}).  Because several interactions have identical $\Lambda_{ij}$ values, these interactions all become antiferromagnetic at exactly the same resolution-parameter value, so the behavior of MRFs only changes for a small number of $\xi$ values. As the rewiring probability $p$ is increased, the degree and $\Lambda_{ij}$ distributions become more heterogeneous, which leads to smoother MRFs. For a rewiring probability of $p=1$, the WS network is just an ER network.

\section{Measuring Distance Between Networks} \label{sec:distmeas}

In the framework that we have introduced in this paper, comparing two networks at the mesoscopic level amounts to characterizing the differences in behavior of the corresponding MRFs. To quantify such differences, we define a distance between two networks with respect to one of the summary statistics as the area between the corresponding MRFs. For example, the distance between two networks $i$ and $j$ with respect to the effective energy $\heffnl$ is given by
\begin{equation}
	d_{ij}^{\hh} = \int_{0}^1{} |\heffnl^{i}(\xi) - \heffnl^{j}(\xi) |\, d\xi\,.
\end{equation}
For the effective entropy and effective number of communities, the distances are given by $d_{ij}^{S} = \int_{0}^{1} |\seff^{i}(\xi) - \seff^{j}(\xi) |\, d\xi$ and $d_{ij}^{\eta} = \int_{0}^{1} |\ceff^{i}(\xi) - \ceff^{j}(\xi) |\, d\xi$, respectively.

We represent the resulting three sets of distances (computed for each pair of networks from the 746 networks that we consider, see Table I) in matrix form as $\mathbf{D}^{\hh}$, $\mathbf{D}^{S}$, and $\mathbf{D}^{\eta}$. These distance measures have several desirable properties. First, they compare MRFs across all network scales (i.e., for all values of $\xi$); second, each distance is bounded between $0$ and $1$; third, the distances are easy to interpret, as each of them corresponds to the geometric area between (a certain dimension of) a pair of MRFs; and finally, we find \emph{a posteriori} that these distances can be used to cluster networks accurately (see the discussions below).

We have computed MRFs for the energy $\mathcal{H}$, entropy $S$, and number of communities $\eta$, but we can proceed similarly with any desired summary statistic. If two diagnostics provide similar information, then one of them can be excluded without significant loss of information. We checked whether the summary statistics were sufficiently different, for the set of networks considered here, for it to be worthwhile to include all of them by calculating the Pearson correlation coefficient between their corresponding distance measures. The correlations between the pairs of distances are $r(d_{ij}^{\hh},d_{ij}^{S})\doteq0.36$, $r(d_{ij}^{\hh},d_{ij}^{\eta})\doteq0.24$, and $r(d_{ij}^{S},d_{ij}^{\eta})\doteq0.58$. These correlations are not sufficiently high to justify excluding any of the summary statistics.

In the interest of parsimony---and given the non-vanishing correlations between the distance measures---we reduce the number of distance measures using principal component analysis (PCA) \cite{pcajolliffe}. Starting with $\mathcal{N}$ networks, we create a $\frac{1}{2}\mathcal{N}(\mathcal{N}-1) \times 3$ matrix in which each column corresponds to the vector representation of the upper triangle of one of the distance matrices $\mathbf{D}^{\hh}$, $\mathbf{D}^{S}$, $\mathbf{D}^{\eta}$, and we perform a PCA on this matrix. We then define a distance matrix $\mathbf{D}^p$ with elements $d^p_{ij} = w_{\hh} d_{ij}^{\hh} + w_{S} d_{ij}^{S} + w_{\eta} d_{ij}^{\eta}$, where the weights are the coefficients for the first principal component, and we normalize the sum of squared coefficients to unity. The coefficients are $w_{\hh}\doteq0.24$, $w_{S}\doteq0.79$, and $w_{\eta}\doteq0.57$. The first component accounts for about 69\% of the variance, so the distances $\mathbf{D}^p$ provide a reasonable single-variable projection of the distances $\mathbf{D}^{\hh}$, $\mathbf{D}^{S}$, and $\mathbf{D}^{\eta}$.

It is important that the distance measures for comparing networks are robust to small perturbations in network structure. Because many of the networks that we study are constructed empirically, they might contain false positives and false negatives. In other words, the networks might falsely identify a relationship where none exists, and they also might fail to identify an existing relationship. Consequently, the topology and edge weights of an observed network might be slightly different than those of the actual underlying network. To test the robustness of our distance measures to such observational errors, we recalculate the MRFs for a subset of relatively small unweighted networks in which, for each network, we rewire a number of edges corresponding to a given percentage of the total number of edges (5\%, 10\%, 20\%, 50\%, or 100\%).  See Appendix~\ref{sec:robust} for more details. (We study networks with up to 1000 nodes and only consider a subset of 25 networks because of the computational costs of rewiring a large number of networks multiple times; however, we have performed the same investigation for 5 different subsets of 25 networks and obtained similar results. We list the networks in each subset in Table~I of the Supplemental Material.) We investigate two rewiring mechanisms: one in which the degree distribution is maintained, where we also ensure after each rewiring that the network forms a single connected component; and another in which the only constraint is that the network continues to consist of a single connected component after each edge rewiring \cite{sneppen}. We find in both cases that the structures of the block-diagonalized distance matrices for the 25 networks (see Figs.\ref{fig:rewiring_percentage} and \ref{fig:full_rewiring_comp} in Appendix \ref{sec:robust}) are robust to random perturbations of the networks, thereby suggesting that our MRF distance measures are not sensitive to small structural perturbations.

\section{Clustering Networks} \label{sec:clustnets}

We assign each of the 746 networks to a category based on its type (see Table~\ref{tbl:category}). Due to the varying availability of different types of network data, the included networks are not evenly distributed across these categories. Many of the networks are either different temporal snapshots of the same system or different realizations of the same type of network. To have a more balanced distribution across the different categories, we focus on 189 of the 746 networks. We only include categories for which we have 8 or more networks, and we selected a subset of networks (uniformly at random) from the larger categories. We also exclude all synthetic networks. See Section IV of the Supplemental Material for the list of networks that we consider and Fig.~1 in Section II of the Supplemental Material for a dendrogram showing a taxonomy we constructed using all 746 networks.

\begin{figure*}[htp]
\begin{center}
\includegraphics[width=1\linewidth]{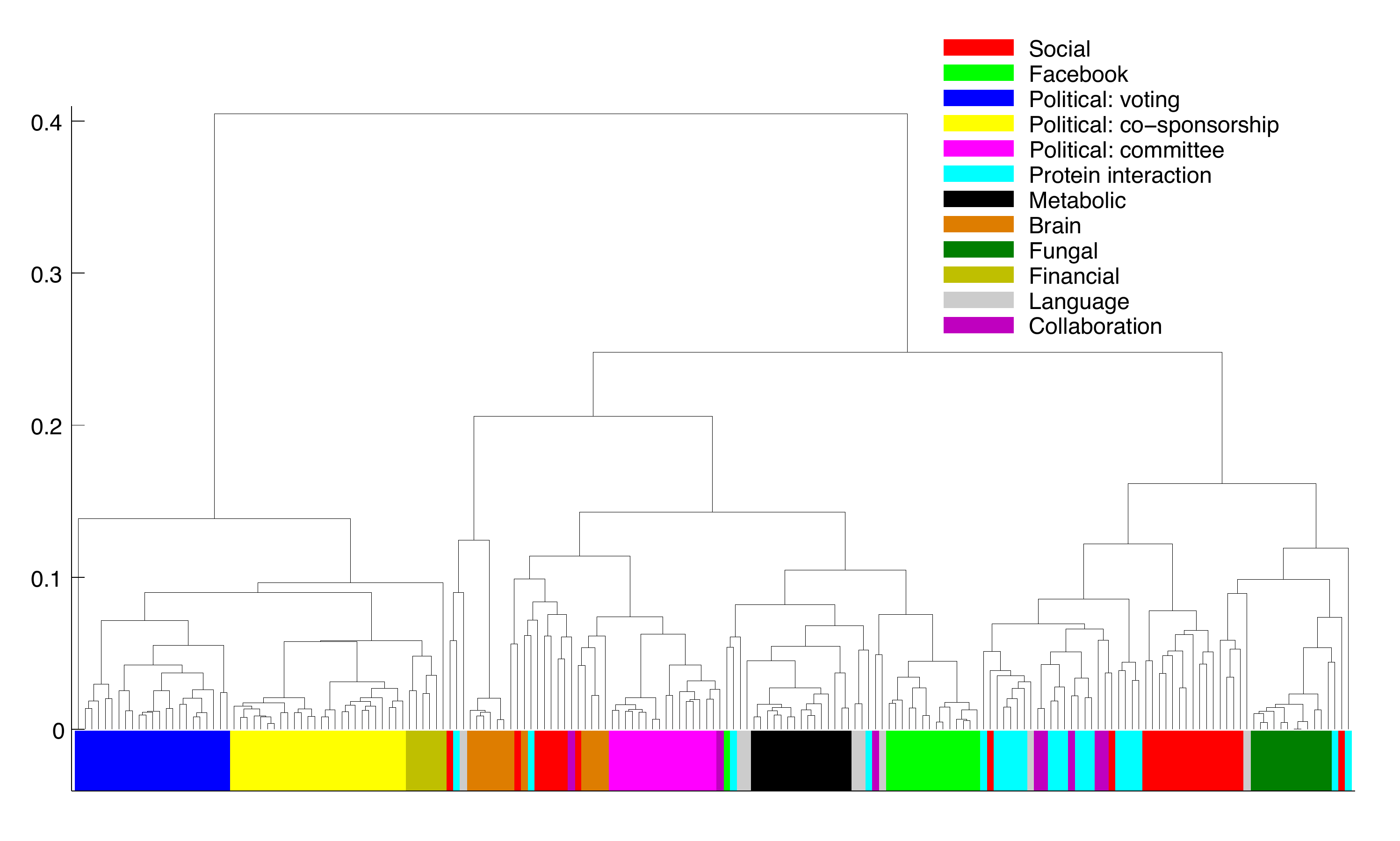}
\end{center}
\caption{(Color online) Taxonomy for 189 networks.  We construct the dendrogram (tree) using the distance $\mathbf{D}^p$ and average linkage clustering. We order the leaves of the dendrogram to minimize the distance between adjacent nodes and color the leaves to indicate the type of network.}
\label{fig:dendrogram_dp}
\end{figure*}

Our primary reason for assigning each network to a category is to use such an external categorization to help assess the quality of taxonomies produced by the unsupervised MRF clustering. For each way of computing distance, we construct a dendrogram for the set of networks using average linkage clustering, which is an agglomerative hierarchical clustering technique \cite{duda,comnotices}\footnote{The three most common linkage clustering algorithms are single, average, and complete linkage clustering; they join clusters of objects based on the smallest, mean, and largest distance between objects in the clusters, respectively. We used the cophenetic correlation $\zeta$ to compare how well each algorithm preserves the pairwise distances between the networks \cite{cophen}. We obtain the following cophenetic correlations for the different linkage clustering algorithms: $\zeta_{\textrm{single}} \doteq 0.65$, $\zeta_{\textrm{avg}} \doteq 0.78$, and $\zeta_{\textrm{comp}} \doteq 0.62$. Hence, dendrograms constructed using average linkage clustering preserve the distances in $\mathbf{D}^p$ better than those constructed using the other two clustering techniques.  We thus use average linkage clustering to construct all of the dendrograms in this paper.}. In Fig.~\ref{fig:dendrogram_dp}, we show a dendrogram obtained from the distance matrix $\mathbf{D}^p$. The colored rectangle underneath each leaf indicates the network category. Contiguous blocks of color demonstrate that networks from the same category have been grouped together using the MRF clustering method, and the presence of such contiguous color blocks is an indication of the success of the MRF clustering scheme.

The assignment of the networks to one of these categories is of course to some extent subjective, as several of the networks could belong to more than one category. For example, we could categorize the network of jazz musicians \cite{jazz} as either a collaboration network or a social network. The initial selection of network categories is also somewhat subjective. One could argue that if one has a social network category, then it is not necessary to have a collaboration network category as well because a collaboration network is a type of social network. We have attempted to maintain a balance between having too many categories and having too few of them. When such ambiguities have arisen, we have systematically chosen the more specific of the relevant categories (e.g., we placed the jazz musician network in the category of collaboration networks rather than in the category of social networks).

\begin{table}
\caption{\label{tbl:category}Network categories, the total number of networks assigned to each category, and the number of networks from each category included in the taxonomy in Fig.~\ref{fig:dendrogram_dp}. For the full taxonomy that uses all 746 networks, see Fig.~1 of the Supplemental Material.}
\begin{center}
\begin{tabular}{|l|c|c|}
\hline
Category	& All networks & Taxonomy networks\\
\hline
\hline
Political: voting	 & 	285 & 23\\
Facebook	& 100 & 15\\
Fungal	& 65 & 12\\
Synthetic	& 58 & 0\\
Financial	& 54 & 6\\
Metabolic	& 43 & 15\\
Social	& 26 & 26\\
Political: cosponsorship	& 26 & 26\\
Other	& 23 & 0\\
Protein interaction	& 22 & 22\\
Political: committee	& 16 & 16\\
Brain	& 12 & 12\\
Language	 & 8 & 8\\
Collaboration & 8 & 8\\
\hline
Total & 746 & 189\\
\hline
\end{tabular}
\end{center}
\label{tbl:cats}
\end{table}

\section{Taxonomies of Empirical Networks}
\label{sec:emptaxons}

All of the networks in some categories appear in blocks of adjacent leaves in the dendrogram in Fig.~\ref{fig:dendrogram_dp}.  For example, there is a cluster of political voting networks at the far left of the dendrogram. This cluster includes voting networks from the US Senate, the US House of Representatives, the UK House of Commons, and the United Nations General Assembly (UNGA). The clustering of these voting networks suggests that there are some common features in the network representations of the different legislative bodies. We also obtain blocks that consist of all political committee networks and all metabolic networks.

There are also several categories for which all except one or two networks cluster into a contiguous block. For example, all but two of the fungal networks appear in the same block and all but one of the Facebook networks are clustered together. The isolated Facebook network is the Caltech network, which is the smallest network of this type and which appears in a group next to that containing all of the other Facebook networks. We remark that the social organization of the community structure of the Caltech Facebook network has been shown to be different from those of the other Facebook networks \cite{traud,datadump}.

Networks of certain categories do not appear in near-contiguous blocks. For example, protein interaction networks appear in several clusters. These networks represent interactions within several different organisms, so we would not expect all of them to be clustered together. Moreover, the data that we employed includes examples of protein interaction networks for the same organism in which the interactions were identified using different experimental techniques, and these networks do not cluster together. This supports previous work suggesting that the properties of protein interaction networks are very sensitive to the experimental procedure used to identify the interactions \cite{lovell,proteinbias}.  Social networks are also distributed throughout the dendrogram. This is unsurprising given the extremely broad nature of the category, which includes networks of very different sizes with edges representing a diverse range of social interactions. The leftmost outlying social network is the network of Marvel comic book characters \cite{marvel}, which is arguably an atypical social network.

The grouping (and, to some extent, the non-grouping) of networks by category suggests that the PCA-distance $\mathbf{D}^p$ between MRFs of different networks produces a sensible taxonomy. It is important to ask, however, whether a simpler approach based on a single network diagnostic, such as edge density, can be comparably successful at constructing a taxonomy. In Appendix~\ref{sec:altchar}, we demonstrate using some well-known diagnostics that this does not appear to be the case, as the diagnostics we tried were unable to reproduce or explain the classifications that we produced using the MRFs.

\begin{figure}[htp]
\begin{center}
\includegraphics[width=1\linewidth]{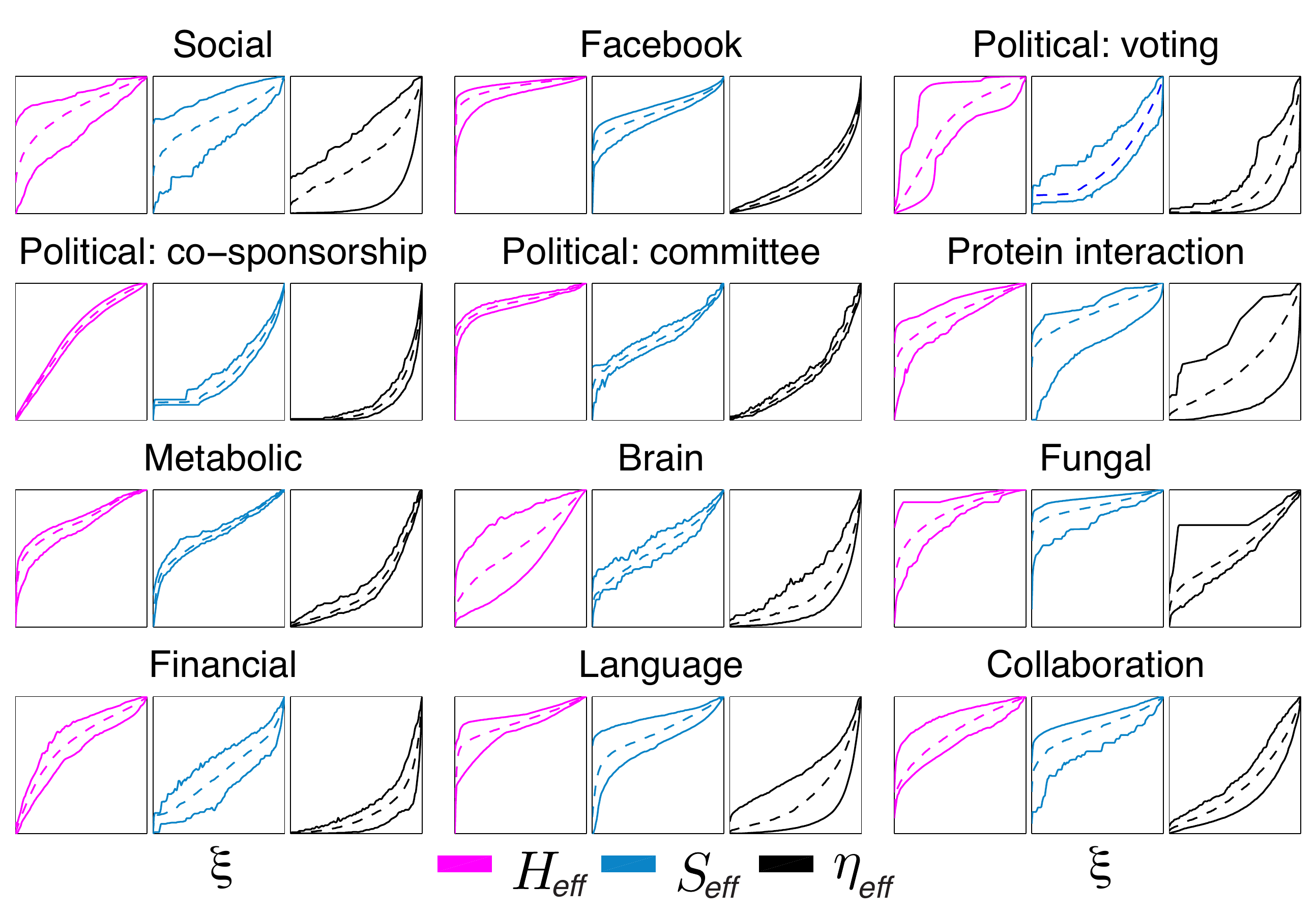}
\caption{(Color online) MRFs for all of the network categories containing at least $8$ networks (see Table~\ref{tbl:cats}). At each value of $\xi$, the upper curve shows the maximum value of $\heffnl$ (pink, left panel in each category), $\seff$ (blue, center panel), and $\ceff$ (black, right panel) for all networks in the category and the lower curve shows the minimum value.  The dashed curves show the corresponding mean MRFs.}
\label{fig:category_MRFs}
\end{center}
\end{figure}

In order to compare the aggregate shapes of the MRFs across categories, we show the bounds of the $\heffnl$, $\seff$, and $\ceff$ curves for each category in Fig.~\ref{fig:category_MRFs}. We again consider all empirical network categories with at least $8$ networks in them. This illustrates that the MRFs for some classes of networks (such as political cosponsorship and metabolic networks) are very similar to each other, whereas there are large variations in the MRFs for other categories (such as social and protein interaction networks). The variety of different MRFs for the social and protein interactions is consistent  with the fact that their constituent networks are scattered throughout the dendrogram in Fig.~\ref{fig:dendrogram_dp}.

\section{Case Studies} \label{sec:casestudies}

We now consider several case studies, in which we generate taxonomies for multiple realizations of particular types of networks and multiple time slices of particular networks. This enables us to compare these networks and (in some cases) illustrate possible connections between network function and mesoscopic network structure.

\subsection{Voting in the United States Senate}

Our first example deals with roll-call voting in the United States Senate \cite{voteview,poole,mccarty,waugh,multislice}. Establishing a taxonomy of networks detailing the voting similarities of individual legislators complements previous studies of these data, and it facilitates the comparison of voting similarity networks across time. We consider Congresses 1--110, which cover the period 1789--2008.  As in Ref.~\cite{waugh}, we construct networks from the roll-call data \cite{poole,voteview} for each two-year Congress such that the adjacency matrix element $A_{ij} \in [0,1]$ represents the number of times Senators $i$ and $j$ voted the same way on a bill (either both in favor of it or both against it) divided by the total number of bills on which both of them voted. Following the approach of Ref.~\cite{poole}, we only consider ``non-unanimous'' roll call votes, which are defined as votes in which at least 3\% of the Senators were in the minority.

\begin{figure}[htp]
\begin{center}
\includegraphics[width=1\linewidth]{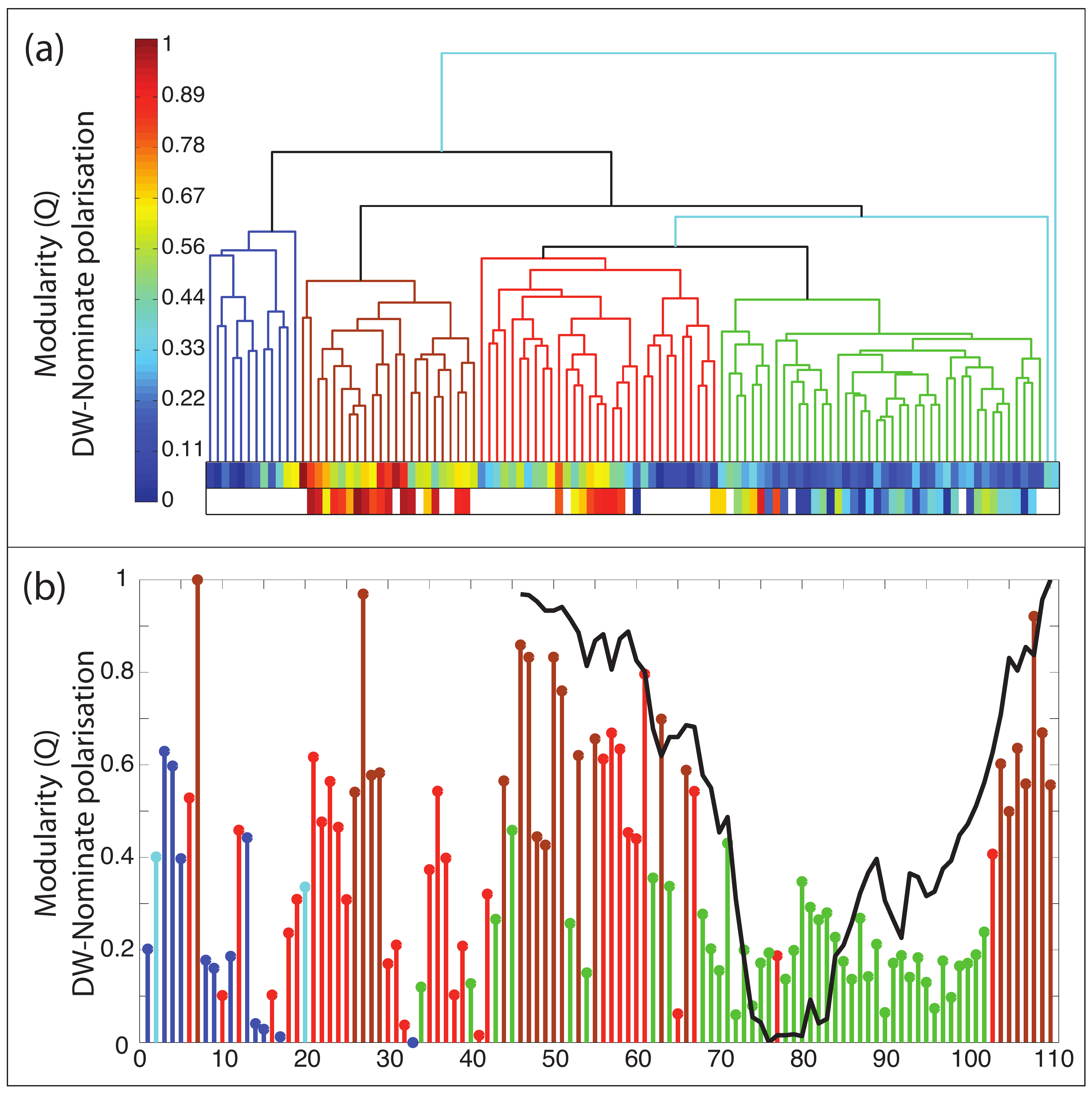}
\caption{(Color online)
\textbf{(a)} Dendrogram for Senate roll-call voting networks for the 1$^{\textrm{st}}$--110$^{\textrm{th}}$ Congresses.  Each leaf in the dendrogram represents a single Senate. Two horizontal color bars below the dendrograms indicate polarization measured in terms of modularity (upper bar) and DW-Nominate scores (lower bar). We color the branches in the dendrogram corresponding to periods of similar polarization.
\textbf{(b)} Polarization of the US Senate as a function of time. The height of each stem indicates the level of polarization measured using modularity, and the color of each stem gives the cluster membership of each Senate in (a). The black curve shows the DW-Nominate polarization. Note that we have rescaled both measures to the interval $[0,1]$.}
\label{fig:senate_dend}
\end{center}
\end{figure}

Much research on the US Congress has been devoted to the ebb and flow of partisan polarization over time and the influence of parties on roll-call voting \cite{waugh,mccarty}. In highly polarized legislatures, representatives tend to vote along party lines, so there are strong similarities in the voting patterns of members of the same party and strong differences between members of different parties. In contrast, during periods of low polarization, the party lines become blurred. The notion of partisan polarization can be used to help understand the taxonomy of Senates in Fig.~\ref{fig:senate_dend}, in which we consider two measures of polarization. The first measure uses DW-Nominate scores (a multi-dimensional scaling technique commonly used in political science \cite{mccarty,poole}), where the extent of polarization is given by the absolute value of the difference between the mean first dimension DW-Nominate scores for members of one party and the same mean for members of the other party \cite{mccarty,poole,voteview}. In particular, we use the simplest such measure of polarization, called MPR polarization, which assumes a competitive two-party system and hence cannot be calculated prior to the $46^{\textrm{th}}$ Senate.  The second measure we consider is network modularity $Q$, which was recently shown to be a good measure of polarization even for Congresses without clear party divisions \cite{waugh}. Modularity is given in terms of the energy $\mathcal{H}$ in Eq.~(\ref{eq:SOM:H}) by $Q = -\mathcal{H}(\lambda=1)/(2m)$. These two measures exhibit fairly close agreement on the level of polarization of each Congress for which they can both be calculated \cite{waugh}.

In Fig.~\ref{fig:senate_dend}(a), we include bars under the dendrograms to represent the two polarization measures, both of which have been normalized to lie in the interval $[0,1]$. The bars demonstrate that Senates with similar levels of polarization (measured in terms of both DW-Nominate scores and modularity values) are usually assigned to the same group, suggesting that our MRF clustering technique groups Senates based on the polarization of roll-call votes. We have also colored dendrogram groups according to their mean levels of polarization using modularity, where the brown group in the dendrogram corresponds to the most highly polarized Senates and the blue group corresponds to the least polarized Senates. Although one ought to expect similarity in the results from the modularity-based measure of polarization and the MRF clustering, it is important to stress that the MRF clustering method is based on different principles; modularity quantifies the extent to which a given network is ``modular'', whereas the MRF clustering explicitly compares the differences in modular structures between any two networks at all scales.

In Fig.~\ref{fig:senate_dend}(a), we also show the clusters that we obtained for the Senate.  They closely match the different periods of polarization that have been identified using modularity and DW-Nominate \cite{waugh}.  The cluster with the highest mean polarization (shown in brown) consists of Senates 7, 26--29, 44, 46-51, 53, 55, 66, and 104--110.  The $104^\textrm{th}$--$110^\textrm{th}$ Congresses correspond to a period of extremely high polarization following the 1994 ``Republican Revolution'', in which the Republican party earned majority status in the House of Representatives for the first time in more than 40 years \cite{mccarty,voteview,waugh}. The cluster with the second highest mean polarization (shown in red) includes several contiguous blocks of Senates, such as those from Congresses 21--25, 35--39, and 56--61.  The $21^\textrm{st}$--$25^\textrm{th}$ Congresses (1829--1839) corresponded to a period of partisan conflict between supporters of John Quincy Adams and Andrew Jackson; it lasted until the emergence of the Whigs and the Democratic party in the $25^\textrm{th}$ Congress \cite{waugh,Kernell-Jacobson-Kousser}.  The American Civil War started during the $37^\textrm{th}$ Congress, and a third party known as the Populist Party was strong during the $56^\textrm{th}$--$58^\textrm{th}$ Congresses.

\begin{figure}[htp]
\begin{center}
\includegraphics[width=1\linewidth]{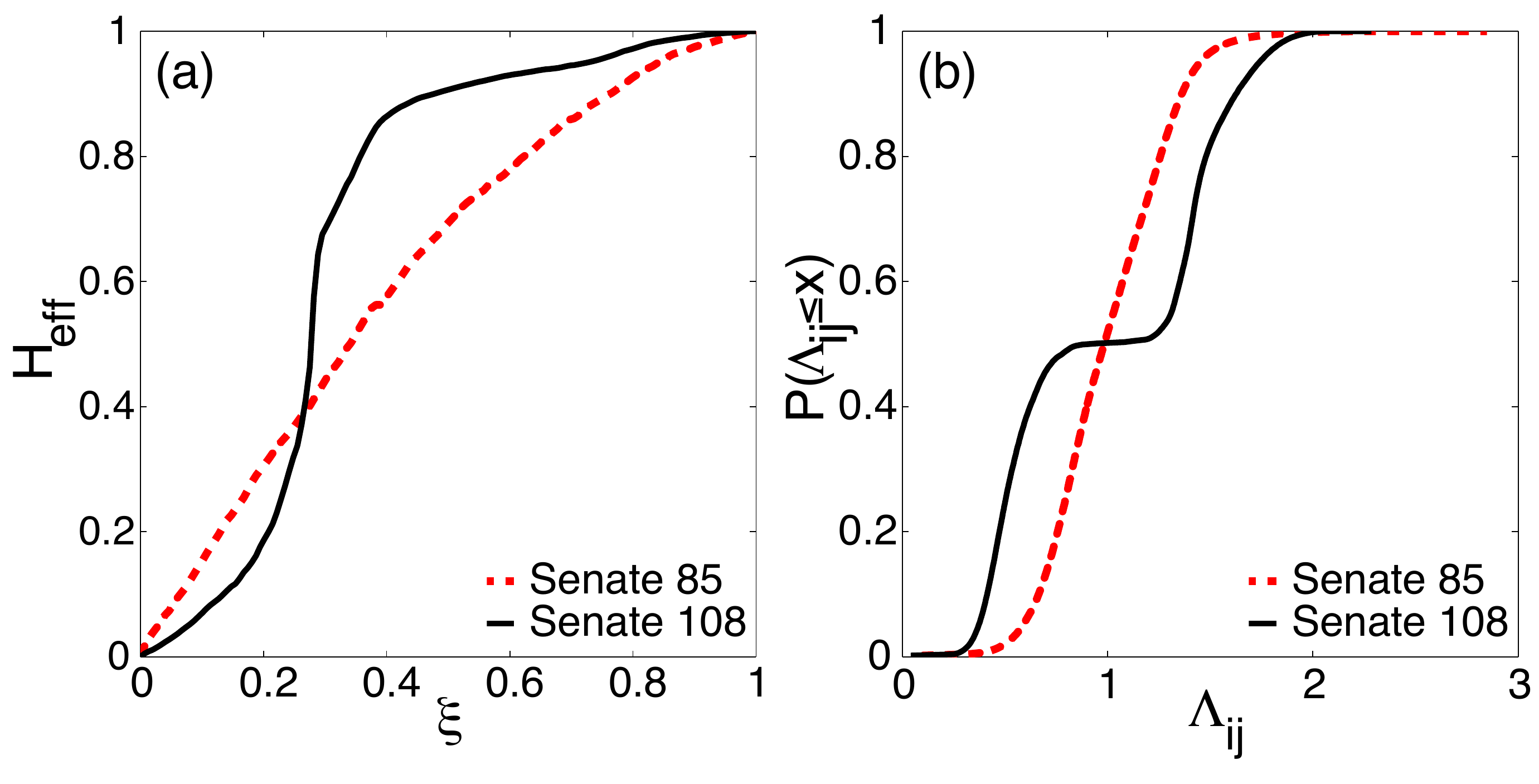}
\end{center}
\caption{(Color online) Comparison of the (low-polarization) $85^\textrm{th}$ Senate and the (high-polarization) $108^\textrm{th}$ Senate. The panels show (\textbf{a}) the $\heffnl$ MRFs and (\textbf{b}) the cumulative distributions of $\Lambda_{ij}$ values.
}
\label{fig:senate_polarization_comp}
\end{figure}

The main differences between different clusters occur in the $\heffnl$ response functions. For the most polarized Senates, there is a sharp shoulder in the $\heffnl$ MRF that becomes less pronounced as the polarization decreases. We illustrate this in Fig.~\ref{fig:senate_polarization_comp}, in which we compare the $\heffnl$ MRFs for the (low-polarization) $85^\textrm{th}$ and (high-polarization) $108^\textrm{th}$ Senates. The shoulder in the $\heffnl$ curve for the $108^\textrm{th}$ Senate is very pronounced, which can be explained by considering the distribution of $\Lambda_{ij}$ values.  The $108^\textrm{th}$ Senate has a bimodal $\Lambda_{ij}$ distribution that contains a trough at $\Lambda_{ij}=1$. Recall that $\Lambda_{ij}=A_{ij}/P_{ij}$, so $\Lambda_{ij}$ compares the observed voting similarity $A_{ij}$ of legislators $i$ and $j$ with the similarity $P_{ij}=k_ik_j/(2m)$ expected from random voting. If $\Lambda_{ij}<1$, legislators $i$ and $j$ vote differently more frequently than expected (with respect to the chosen null model); if $\Lambda_{ij}>1$, they vote more similarly than expected. Therefore, the peaks in the $\Lambda_{ij}$ distribution above and below $1$ correspond, respectively, to intra-party and inter-party voting blocs. In a Senate with low polarization, legislators from different parties often vote in the same manner, so the values of $\Lambda_{ij}$ no longer separate two distinct types of behavior.

We also examined roll-call voting networks in the US House of Representatives and found many similar features as the ones that we have presented for the US Senate. For example, the highly polarized $104^\textrm{th}$--$110^\textrm{th}$ Congresses, which followed the ``Republican Revolution'', appear in the same cluster for both the House and Senate. We also observed some differences in the clusters for the two chambers. For example, the $78^\textrm{th}$--$102^\textrm{nd}$ Senates all appeared in the same cluster. For the House, however, Congresses 80, 88, 89, and 98--102 did not appear in the same cluster as the other Congresses between 78 and 102; instead, they appeared in a cluster that also included the $26^\textrm{th}$--$28^\textrm{th}$ Houses. This was a particularly eventful period: the $25^\textrm{th}$ Congress saw the emergence of the Whigs and the Democratic Party, and the abolitionist movement was also prevalent (e.g., the Amistad seizure occurred in 1839 during the $26^\textrm{th}$ Congress).

\subsection{Voting in the United Nations General Assembly}

The United Nations General Assembly (UNGA) is one of the principal organs of the United Nations (UN), and it is the only part of the UN in which all member nations have equal representation. Although most resolutions are neither legally nor practically enforceable because the General Assembly lacks enforcement powers on most issues, it is the only forum in which a large number of states meet and vote regularly on international issues.  It also provides an interesting point of comparison with roll-call voting in the US Congress, as the level of agreement on UN resolutions tends to be much higher than that in the Senate and House \cite{macon}.

We study voting for the 1$^{\textrm{st}}$--63$^{\textrm{rd}}$ sessions (covering the period 1946--2008), where each session corresponds to a year \footnote{We excluded the 19$^{\textrm{th}}$ session because only one resolution was considered by the UNGA that year.}.  For each session, we define an adjacency matrix $\mathbf{A}$ whose elements $A_{ij}$ represent the number of times countries $i$ and $j$ voted in the same manner in a session (i.e., the sum of the number of times both countries voted yea on the same resolution, both countries voted nay on the same resolution, or both countries abstained from voting on the same resolution) divided by the total number of resolutions on which the UNGA voted in a session. The matrix $\mathbf{A}$, with elements $A_{ij}\in[0,1]$, thereby represents a (similarity) network of weighted edges between countries.

\begin{figure}[htp]
\begin{center}
\includegraphics[width=1\linewidth]{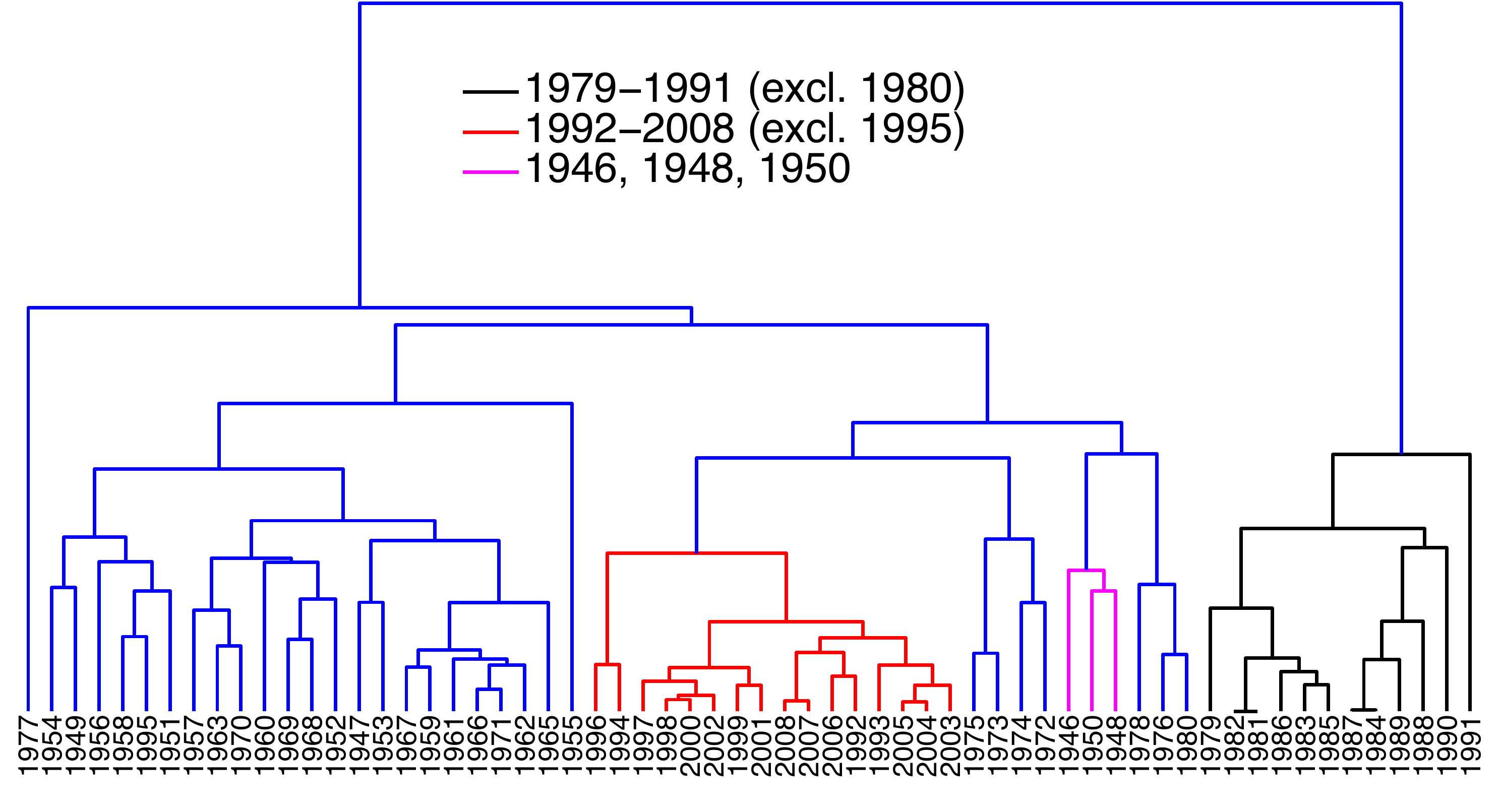}
\caption{(Color online) Dendrogram for the United Nations General Assembly resolution voting network for the 1$^{\textrm{st}}$--63$^{\textrm{rd}}$ sessions (excluding the 19$^{\textrm{th}}$ session), covering the period 1946--2008.  Each leaf in the dendrogram represents a single session.  In the text, we discuss the coloring of groups of branches in the dendrogram.}
\label{fig:UN_dend}
\end{center}
\end{figure}

We cluster UNGA sessions by comparing MRFs for the corresponding voting networks. In Fig.~\ref{fig:UN_dend}, we plot a dendrogram of the UNGA sessions and highlight some of the clusters, which correspond to notable periods in the recent history of international relations.  The red cluster in the middle of the dendrogram consists of all post-Cold War sessions (1992--2008) except 1995. This group forms a larger cluster with some UNGA sessions from the 1970s and a cluster consisting of 1946, 1948, and 1950. These three sessions (shown in magenta) are all noteworthy: 1946 was the first session of the UNGA, the Universal Declaration of Human Rights was introduced during the 1948 session, and the ``Uniting for Peace'' resolution was passed during the 1950 session.  At the rightmost part of the dendrogram, we color in black a group that consists of all sessions from 1979 to 1991 (excluding 1980). The beginning of this period marked the end of D\'{e}tente between the Soviet Union and the US following the former's invasion of Afghanistan at the end of 1979, and the end of this period saw the end of the Cold War.  The large blue cluster in the leftmost part of the dendrogram consists primarily of sessions from before 1971 (though it also includes the sessions in 1977 and 1995).

\subsection{Facebook}

We now consider Facebook networks for 100 US universities \cite{traud,datadump}.  The nodes in each network represent users of the Facebook social networking site, and the unweighted edges represent reciprocated ``friendships'' between users at a single-time snapshot in September 2005. We consider only edges between students at the same university, as this allows us to compare the structure of the networks at the different institutions. These networks represent complete data sets obtained directly from Facebook. In contrast to the previous examples, we are not comparing snapshots of the same network at different times but are instead comparing multiple realizations of the same type of network that have evolved independently.  Such real-world ensembles of network data are rare, and constructing a taxonomy will hopefully allow us to compare and contrast the social organization at these institutions.

\begin{figure}[htp]
\begin{center}
\includegraphics[width=1\linewidth]{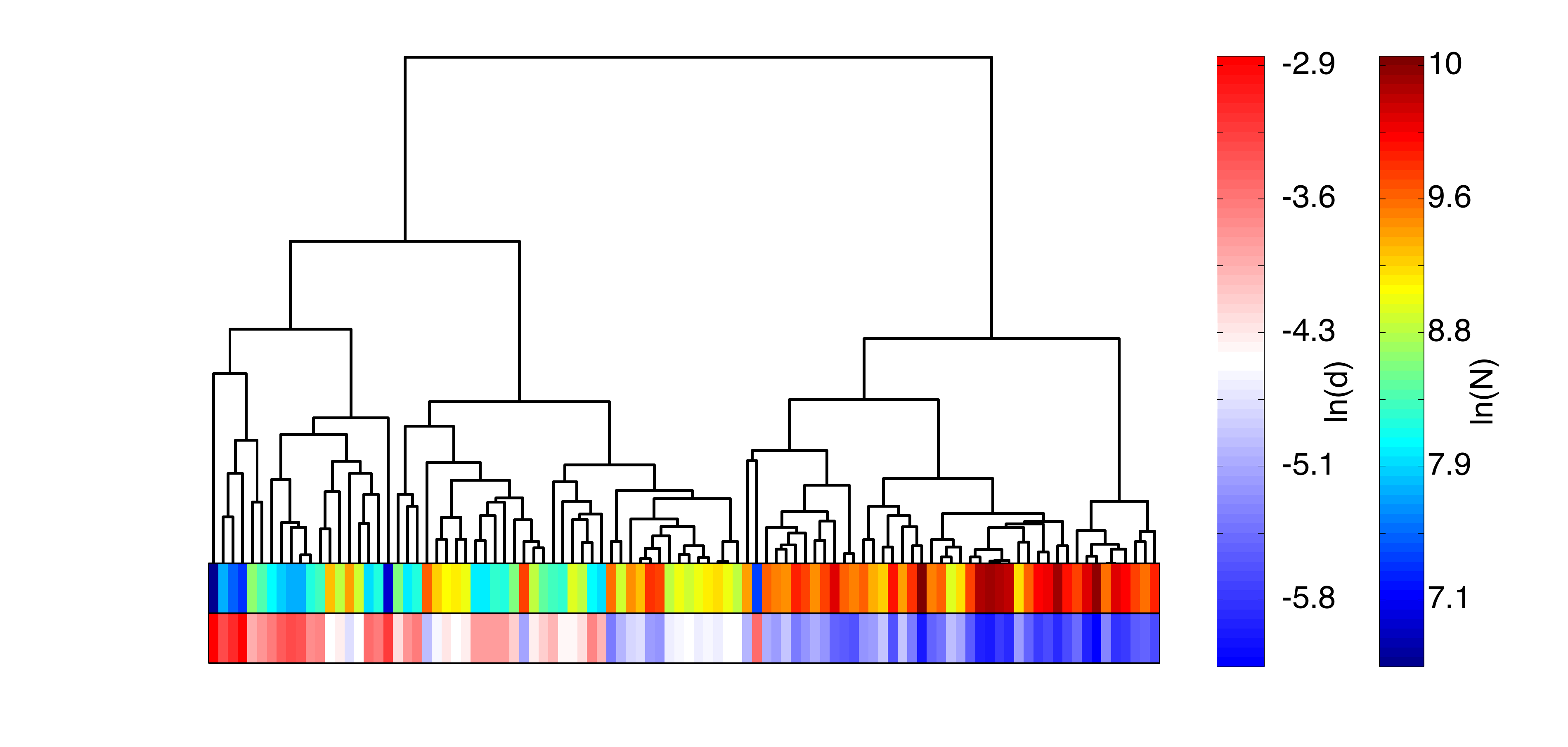}
\end{center}
\caption{(Color online) Dendrogram for 100 Facebook networks of US universities at a single-time snapshot in September 2005. We order the leaves of the dendrogram to minimize the distance between adjacent nodes. The color bars below the dendrogram indicate (top) the number of nodes in the networks $N$ and (bottom) the fraction of possible edges that are present $d$.}
\label{fig:fb_dend}
\end{figure}

In Fig.~\ref{fig:fb_dend}, we show the dendrogram for Facebook networks that we produced by comparing MRFs. The two color bars below the dendrogram indicate (top) the number of nodes $N$ in each network and (bottom) the fraction of possible edges $d$ that are present (i.e., edge density). The Facebook networks range in size from 762 to 41,536 nodes, and the edge density varies from 0.2\% to 6\%. In contrast to previous examples, we observe in this case that two simple network properties appear to explain most of the observed clustering of the networks. An important feature of this example is that the $\heffnl$, $\seff$, and $\ceff$ MRFs are each very similar in shape and lie in a narrow range across all 100 institutions (see Fig.~\ref{fig:category_MRFs}). Such extreme similarity is remarkable---as one can see in Fig.~\ref{fig:category_MRFs}, this contrasts starkly with most of the other examples---and it suggests that all of the Facebook networks have very similar mesoscopic structural features. If one also considers demographic information, then one can find interesting differences between the networks \cite{traud,datadump}, but the structural similarity is striking.

\subsection{Fungi}

We also examined fungal mycelial networks extracted from time series of digitized images of colony growth. In these undirected, planar, weighted networks, the nodes represent hyphal tips, branch points, or anastomoses (hyphal fusions), and the edges represent the interconnecting hyphal cords weighted by their conductivity \cite{fricker,fricker3,fricker4}.  For comparison, we also digitized weighted networks of the acellular slime mold \emph{Physarum polycephalum} \cite{fricker2}.  Fungal networks look like trees but contain additional edges (known as \emph{cross-links}) that generate cycles.

\begin{figure}[htp]
\begin{center}
\includegraphics[width=1\linewidth]{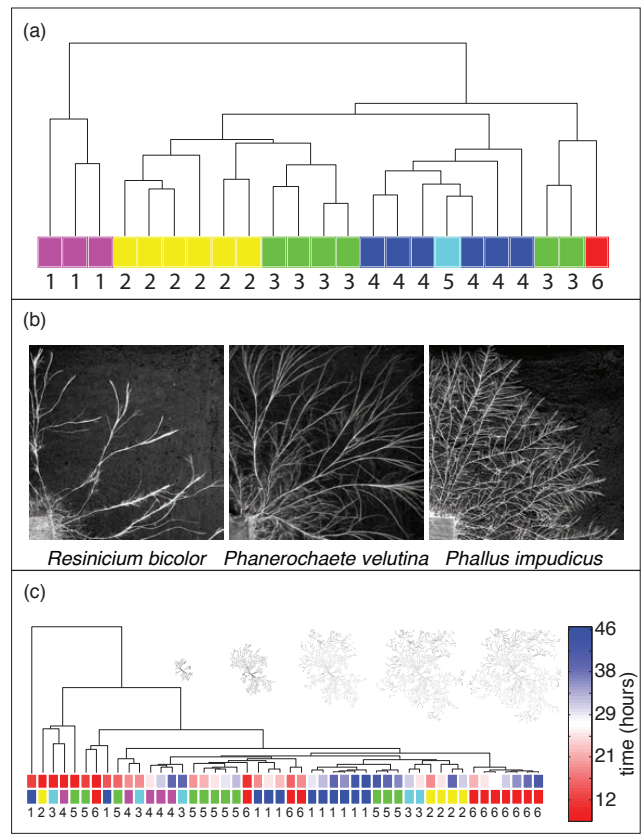}
\caption{(Color online) (a) Dendrogram of networks for six different species of \emph{Saprotrophic basidiomycetes} and the slime mold \emph{Physarum polycephalum}. Each leaf represents a replicate experiment.  The colors and numbers correspond to the species as follows: (1) \emph{Resinicium bicolor}, (2) \emph{Physarum polycephalum}, (3) \emph{Phallus impudicus}, (4) \emph{Phanerochaete velutina}, (5) \emph{Stropharia caerulea}, and (6) \emph{Agrocybe gibberosa}. (b) Images illustrating the network structure of the different species \cite{fricker3}. (c) Dendrogram of network development in six replicate time series of \emph{Phanerochaete velutina}. We color the leaves by time, and the color bar underneath the leaves indicates experiment number ($1,\ldots,6$).  In the inset, we show extracted networks that illustrate the transition from simple branching trees to increasing levels of interconnection (i.e., cross-linking) with time.
}
\label{fig:fungus}
\end{center}
\end{figure}

As shown in Fig.~\ref{fig:fungus}(a), we find using our method that replicate networks from different species at comparable time points are grouped together. Furthermore, the aggregate clustering pattern reflects increasing levels of cross-linking that are characteristic of different species, as illustrated in Fig.~\ref{fig:fungus}(b); this ranges from the low levels in \emph{Resinicium bicolor} to intermediate levels in \emph{Phanerochaete velutina} and highly cross-linked networks formed by \emph{Phallus impudicus}.  By constructing a dendrogram for only one species but including data from repeated experiments and over time (see Fig.~\ref{fig:fungus}(c)), we observe a progression from trees at early developmental times to an increasingly cross-linked network later in mycelium growth \cite{fricker,heaton}. In early growth, the developmental stage appears to dominate the clustering pattern, as networks from different replicates but of similar age are grouped together. At later times, however, networks show a high aggregate level of similarity, and the fine-grained clustering predominantly reflects the subtle changes in structure evolving within each replicate.

\subsection{New York Stock Exchange}
\label{subsec:nyse}

As our final example, we consider a set of stock-return correlation networks for the New York Stock Exchange (NYSE), which is the largest stock exchange in the world (as measured by the aggregate US dollar value of the securities listed on it). Each node represents a stock, and the strength of the edge connecting stocks $i$ and $j$ is linear in the Pearson product-moment correlation coefficient between the daily logarithmic returns of the stocks \cite{nyse}. We consider $N=100$ stocks during the time period 1985--2008 and construct a network for each 6 months of data. This yields a sequence of fully-connected, weighted adjacency matrices whose elements quantify the similarity of two stocks (normalized to the unit interval for each time window).

\begin{figure}[htp]
\begin{center}
\includegraphics[width=1\linewidth]{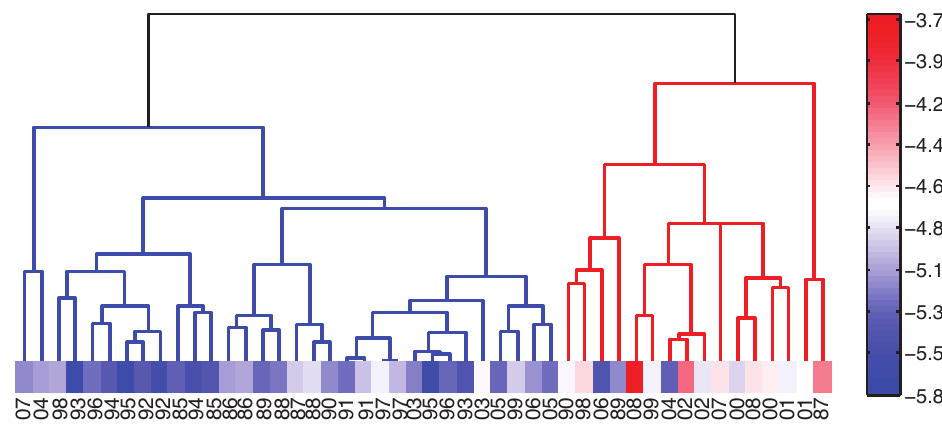}
\caption{(Color online) Dendrogram for 48 NYSE networks during the period 1985--2008 \cite{nyse}. Observe the clear split of the dendrogram into two clusters (a blue group on the left and a red group on the right).  Leaf color indicates mean daily volatility of the composite index.}
\label{stock}
\end{center}
\end{figure}

We show the dendrogram for the NYSE networks in Fig.~\ref{stock}. The first division of these networks classifies them into two groups (which we have colored in blue and red). The red cluster appears to correspond to periods of market turmoil, including the networks for the second half of 1987 (including the Black Monday crash of October 1987), all of 2000--2002 (including and following the bursting of the dot-com bubble), and the second half of 2007 and all of 2008 (including the recent credit and liquidity crisis). The value of the NYSE composite index, which measures the aggregate performance of all common stocks listed on the NYSE \cite{nysecompindex}, supports our hypothesis that the red cluster is associated with periods of market turmoil. Indeed, the networks in the red cluster correspond (with one or two exceptions) to the periods of high volatility of the composite index (see Fig.~\ref{stock}).

\section{Conclusions} \label{sec:end}

We have developed an approach that facilitates the comparison of diverse networks by summarizing network community structure using what we call mesoscopic response functions (MRFs). We have demonstrated how this approach can be used to group networks both across categories and within categories. Our work builds on prior research on network community structure, which has focused predominantly on algorithmic detection of the communities rather than on subsequently using the communities for applications (such as comparing sets of networks).

The development of algorithmic methods to detect communities is frequently motivated by the idea that the community structure of a network representing a system has some bearing on the function of the system. If different networks perform different functions---and if their functions are constrained, at least in part, by their mesoscopic structure---then it should be possible in principle to derive a functional classification of networks based on community structure. Although this has mostly been presented as a presumption in the existing literature, it is actually an empirically testable hypothesis. Indeed, we have shown in the present paper that one can systematically exploit mesoscopic structure to obtain useful comparisons of networks. This allows one to derive taxonomies for networks that also appear to have correspondence with functional similarities. We observed that networks that were not grouped with other members of the same class appeared to be unusual in some respects, and we also demonstrated that we could detect historically-noted financial and political changes from time-ordered sequences of networks.

We believe that our framework has the potential to aid in the exploration and exploitation of similarities in network structures across both network types and disciplinary boundaries.

\section*{Acknowledgments}

Each of the networks that we have used is described in detail in the Supplemental Material, where we have also provide relevant citations for the data sources.  We are particularly grateful to A.~D'Angelo and Facebook, A.~Merdzanovic, M.~E.~J.~Newman, E.`Voeten, and Voteview \cite{voteview} (maintained by K.~Poole and H.~Rosenthal) for data. We thank A.~Lewis, G.~Villar, S.~Agarwal, D.~Smith, and the anonymous referees for useful suggestions.  We adapted code originally written by R.~Guimer\`a for the simulated annealing algorithm and code by R.~Lambiotte for the Louvain algorithm. This research was supported in part by the Fulbright Program (JPO) and NIH grant P01 AG-031093 (JPO), the James S.~McDonnell Foundation (MAP: \#220020177), NSF (PJM: DMS-0645369), and EPSRC and BBSRC (NSJ: EP/I005765/1, EP/I005986/1, and BBD0201901).

\appendix

\section{Robustness of Clustering} \label{sec:robust}
To examine the robustness of our clustering to false positives (false links) and false negatives (false non-links), we consider two network rewiring mechanisms, and we apply the rewiring to each network in a subset of 25 networks highlighted in Table 2 of the Supplemental Material. The first step in the procedure is to randomly rewire a number of edges corresponding to a given percentage (5\%, 10\%, 20\%, 50\%, or 100\%) of the total number of edges in the network, subject to the constraints that we preserve the networks's degree distribution and the fact that it consists of a single connected component \cite{maslov}.  (That is, such a rewiring of a number of edges equal to x\% of the $L$ edges in a network means that we perform $\lceil xL \rceil$ rewiring steps;  the same edge can be rewired multiple times.)  Second, we randomly rewire a given number of the edges subject only to the constraint that we the rewired network still consists of a single component. 

Because we are perturbing the original network, we focus on the distance matrices $\mathbf{D}^{\hh}$, $\mathbf{D}^{S}$, and $\mathbf{D}^{\eta}$ as they can be calculated directly for each network. We consider 25 of the 746 original networks of varying sizes and edge densities; we highlight these networks in bold in Table~II of the Supplemental Material. In Fig.~\ref{fig:rewiring_percentage}, we show the distance matrices for this subset of networks when different percentages of edges have been rewired with the degree distribution preserved. The first column shows the matrices for the original networks. (Note that the node orderings for $\mathbf{D}^{\hh}$, $\mathbf{D}^{S}$, and $\mathbf{D}^{\eta}$ are not necessarily the same in Fig.~\ref{fig:rewiring_percentage} because of the block-diagonalization of matrices.) The subsequent columns show the mean distance matrices as increasing numbers of edges are rewired; for a given row, the node ordering in each column is fixed. The distance matrices for the randomizations are the mean pairwise distances between networks, where the mean is calculated over all possible pairs between 10 perturbations of each network. More precisely, let $A$ and $B$ represent two different (unperturbed) networks and let the sequences $A_1, A_2, \ldots, A_{10}$ and $B_1, B_2, \ldots, B_{10}$ represent 10 realizations of the perturbation process (e.g., at the 5\% level) for the networks. To calculate the distance between $A$ and $B$ under perturbation, we find for each $j \in \{1,\ldots, 10\}$ the distances between $A_j$ and $B_1$, $A_j$ and $B_2$, \ldots, and $A_j$ and $B_{10}$. We then calculate the mean of the ensuing $10 \times 10 = 100$ distance values. Based on visual inspection of Fig.~\ref{fig:rewiring_percentage}, the matrices for the first few columns for all of the distances are fairly similar to the original distance matrices. This suggests some notion of robustness in our clustering technique. We study only 25 networks because of the computational costs of rewiring a large number of networks multiple times; however, we have performed the same investigation for 5 different subsets of 25 networks and obtained similar results. We list the networks in each subset of 25 in Table~I in the Supplemental Material.

\begin{figure}[tbp]
\begin{center}
\includegraphics[width=1\linewidth]{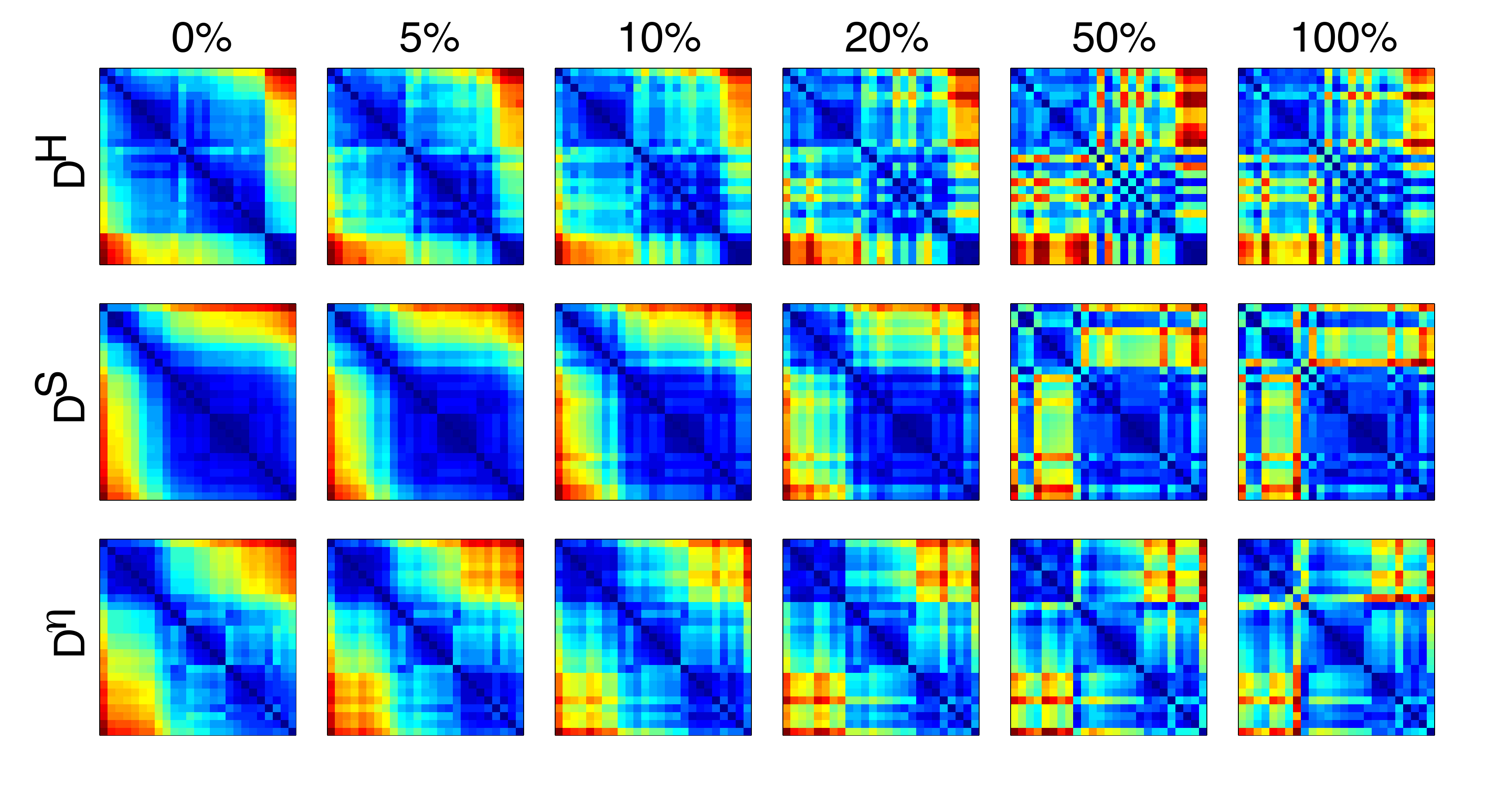}
\end{center}
\caption{(Color online)
Block-diagonalized mean distance matrices $\mathbf{D}^{\hh}$ (top row), $\mathbf{D}^{S}$ (middle row), and $\mathbf{D}^{\eta}$ (bottom row) for the 25 networks listed in bold in Table~II of the Supplemental Material. The columns show the mean-distance matrices following randomizations of the original network in which a given percentage of edges are rewired and the degree distributions of the networks are preserved.  (We also constrain each rewired network to consist of a single connected component.)  The ordering of the nodes in the plots is fixed for each row.  The first column shows the distance matrix for the original networks.  The distance matrices for the randomizations are the mean pairwise distances between networks.}
\label{fig:rewiring_percentage}
\end{figure}

\begin{figure}[tbp]
\begin{center}
\includegraphics[width=0.88\linewidth]{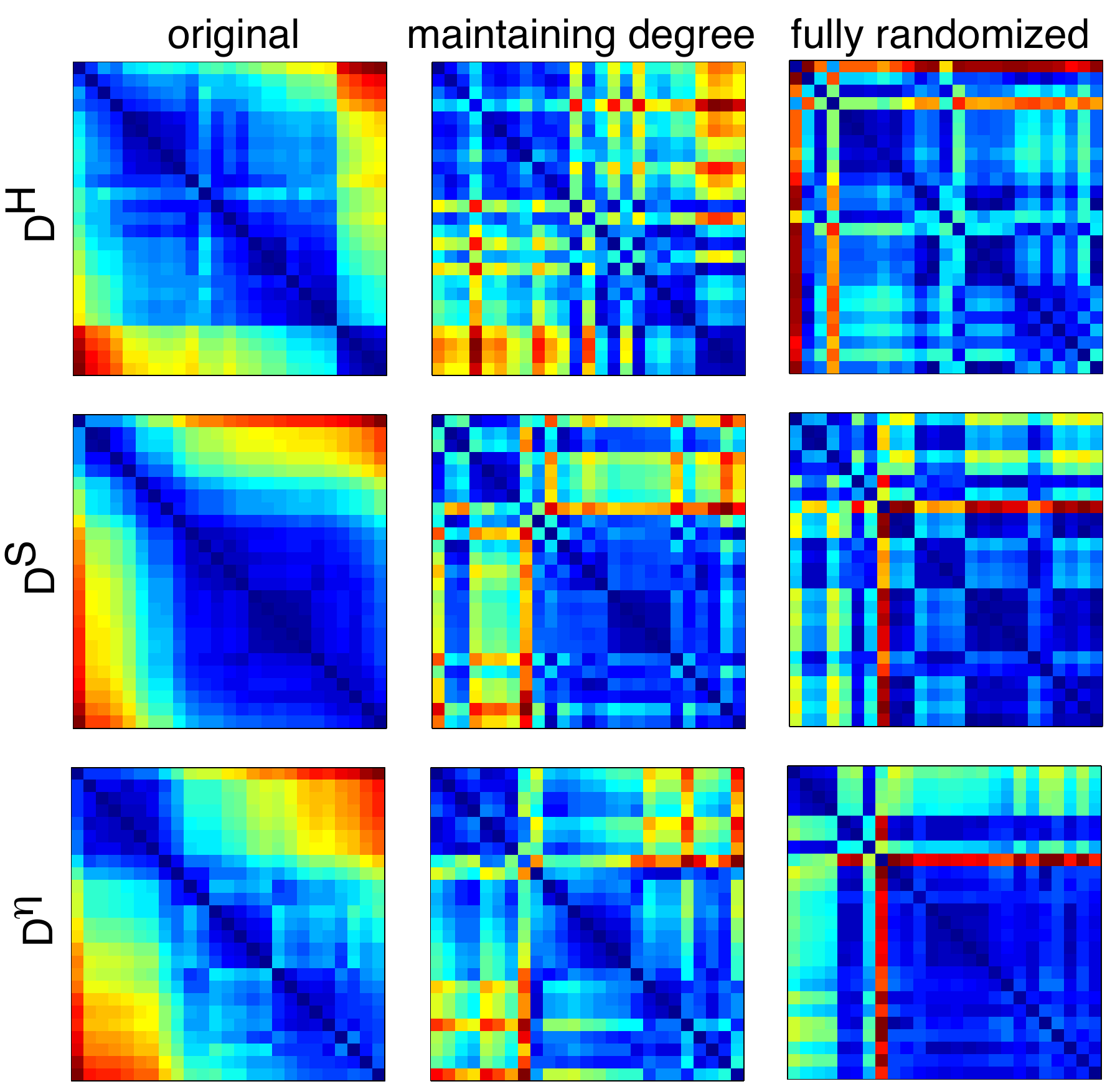}
\end{center}
\caption{(Color online)
Block-diagonalized distance matrices $\mathbf{D}^{\hh}$ (top row), $\mathbf{D}^{S}$ (middle row), and $\mathbf{D}^{\eta}$ (bottom row) for the 25 networks listed in bold in Table II of the Supplemental Material. The first column shows the distance matrices for the original networks. The second column shows the mean distance matrices following randomizations of the original networks in which 10 times the total number of edges in the networks have been rewired such that the degree distributions are preserved and the rewired networks each consist of a single connected component.  The third column shows the mean distance matrices following randomizations of the original networks in which 10 times the total number of edges in the networks have been rewired but only the fact that the networks consist of single connected components is preserved (i.e., the degree distributions are not preserved). The distance matrices for the randomizations are composed of the mean pairwise distances between the networks.}
\label{fig:full_rewiring_comp}
\end{figure}

To carry out a more thorough randomization of each network, we now rewire every edge in the network 10 times on average. In Fig.~\ref{fig:full_rewiring_comp}, we show the $\mathbf{D}^{\hh}$, $\mathbf{D}^{S}$, and $\mathbf{D}^{\eta}$ mean-distance matrices for this number of rewirings. We again calculate the mean distance using the method described in the previous paragraph. The first column again shows the distance matrices for the original networks. The second and third columns show the distance matrices for randomizations in which the degree distribution is preserved and destroyed, respectively. The node orderings of the matrices in the second and third columns are again the same as the orderings for the matrix of the first column of the corresponding row. The second column in Fig.~\ref{fig:full_rewiring_comp} demonstrates that some block structure remains in the distance matrices when the degree distribution is preserved. The third column shows that much of this structure is destroyed (though some block structure is still visible) when the degree distribution is not preserved. When the networks are ``fully randomized'' in this way---with the only constraint being that each rewired network must consist of a single connected component---one is in effect producing random graphs. These random graphs might, however, still have some common properties, such as the number of nodes and the edge density.

\section{Computational Heuristics}\label{sec:heuristics}

\subsection{Robustness of Network MRFs}

We detected all communities in the main text using the locally greedy Louvain algorithm \cite{blondel}; however, several alternative heuristics exist, so we now investigate whether the choice of heuristic has any effect on the results. In Ref.~\cite{modcaution}, Good et al.~demonstrated that there can be extreme near-degeneracies in the energy function, in particular an exponential number of low-energy (i.e., high-modularity) solutions. Given this, it is unsurprising that different energy-optimization heuristics can yield very different partitions for the same network. Good et al.~suggested that the reason for this behavior is that different heuristics sample different regions of the energy landscape. Because of the potential sensitivity of results to the choice of heuristic, one should treat individual partitions by particular heuristics with caution. However, one can have more confidence in the validity of the partitions if different heuristics produce similar results. Here we compare the results for the Louvain algorithm \cite{blondel} with those for a spectral algorithm \cite{newmanpnas2006} and simulated annealing \cite{simulatedannealing}.

\begin{figure}[htp]
\begin{center}
\includegraphics[width=1\linewidth]{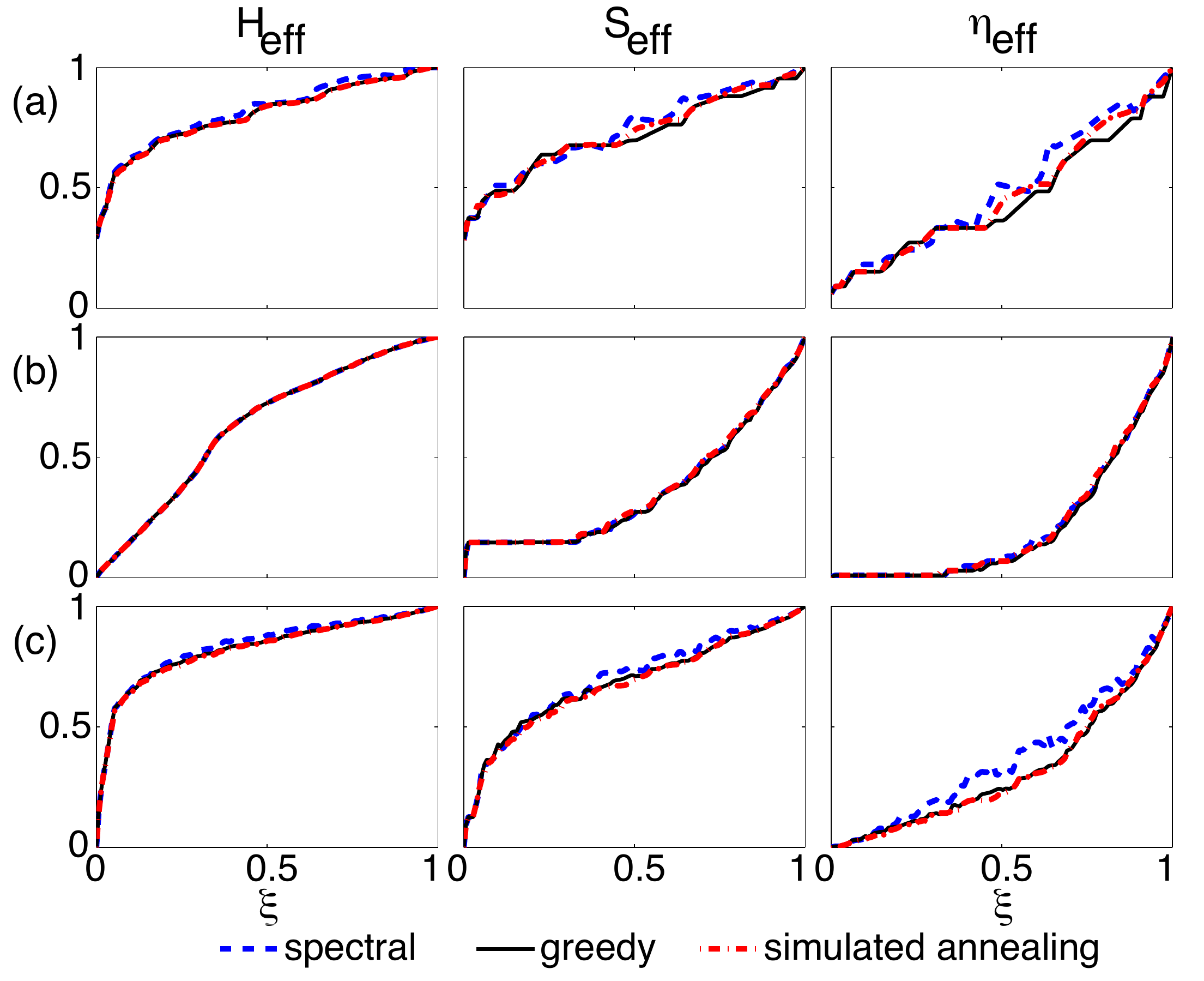}
\end{center}
\caption{(Color online)
Comparison of the MRFs produced using spectral \cite{newmanpnas2006}, Louvain \cite{blondel}, and simulated annealing \cite{simulatedannealing} optimization heuristics. We show the MRFs for the (\textbf{a}) Zachary Karate Club network \cite{karate}, (\textbf{b}) the roll-call voting network of the 102$^{\mbox{nd}}$ US Senate \cite{waugh,poole,mccarty}, and (\textbf{c}) the Garfield small-world citations network \cite{garfield}.}
\label{fig:algo_comp_MRFs}
\end{figure}

In Fig.~\ref{fig:algo_comp_MRFs}, we show MRFs for three networks calculated using Louvain \cite{blondel}, spectral \cite{newmanpnas2006} and simulated annealing algorithms \cite{simulatedannealing}. For all three networks, the three algorithms agree very closely on the shapes of the $\mathcal{H}$, $S$, and $\eta$ MRFs. The MRFs are most similar for the roll-call voting network of the 102$^{\mbox{nd}}$ US Senate \cite{waugh,poole,mccarty}, and the $\mathcal{H}$ MRF is almost identical for the three heuristics. In general, we observe the largest differences in the shapes of the MRFs when using the spectral algorithm. The spectral algorithm that we used begins by finding a partition of the network into exactly two components such that the energy is minimized (among all bipartitions). It then recursively partitions the smaller networks into two groups until no decrease in energy can be obtained through bipartitioning. At each step, this algorithm only finds the optimal partition of each community into two smaller communities even though a split into more communities could yield a lower energy. Given this, it is unsurprising that the spectral algorithm often identifies partitions further from the optimum than the other heuristics. For the remainder of this section, we therefore only compare the Louvain and simulated annealing algorithms.

\subsection{Robustness of Resulting Network Taxonomies}

Although Fig.~\ref{fig:algo_comp_MRFs} shows good agreement between the shapes of the MRFs that we obtain from the different computational heuristics, we nevertheless check that the small differences that do occur do not have a significant effect on the resulting network taxonomy. Because of the computational cost of detecting communities using simulated annealing, we investigate the effect on the taxonomy using a subset of small networks. We highlight all of the networks that we consider with an asterisk $(^*)$ in Table~II of the Supplemental Material. (The largest network that we include is the cat brain cortical/thalmic network \cite{cat_brain}, which has 1,170 nodes.). Indeed, MRFs for small networks tend to be much noisier than those for large networks---see, for example, Fig.~\ref{fig:algo_comp_MRFs}(a), which shows the MRFs for the 34-node Zachary Karate Club network---so we expect that any differences between algorithms are likely to be more pronounced for small networks.

\begin{figure}[htp]
\begin{center}
\includegraphics[width=0.90\linewidth]{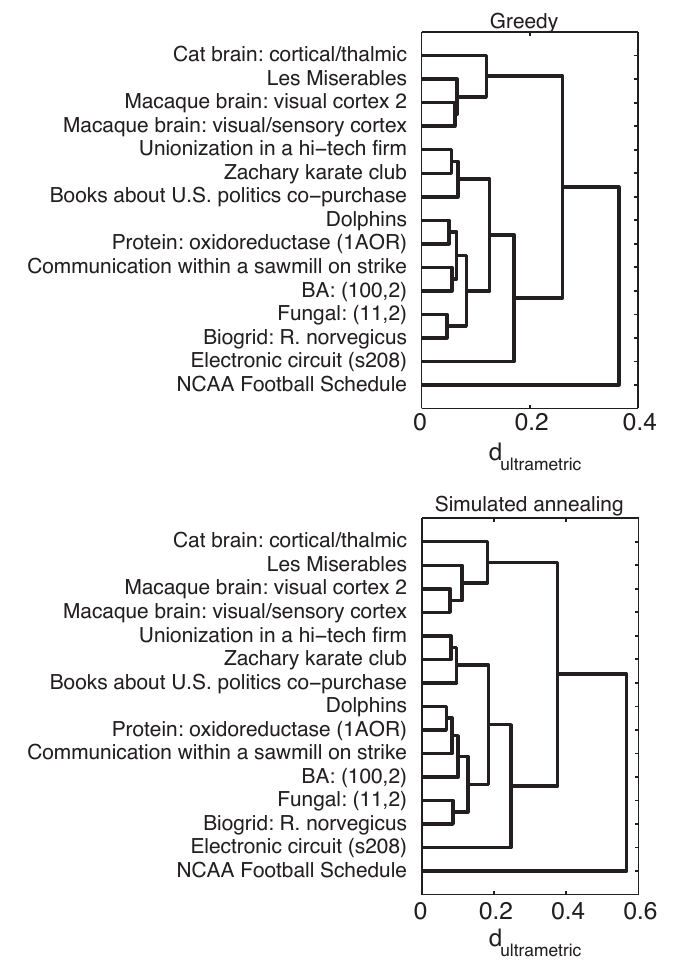}
\end{center}
\caption{(Color online)
Comparison of the dendrograms produced using a Louvain algorithm (top panel) and simulated annealing (bottom panel) for a subset of 15 networks. The only difference between the two dendrograms is the order in which the ``Communication within a sawmill on strike" and the ``BA: (100,2)" networks cluster and the distances at which the other networks cluster.}
\label{fig:algo_comp_dend}
\end{figure}

\begin{figure}[tbp]
\begin{center}
\includegraphics[width=1\linewidth]{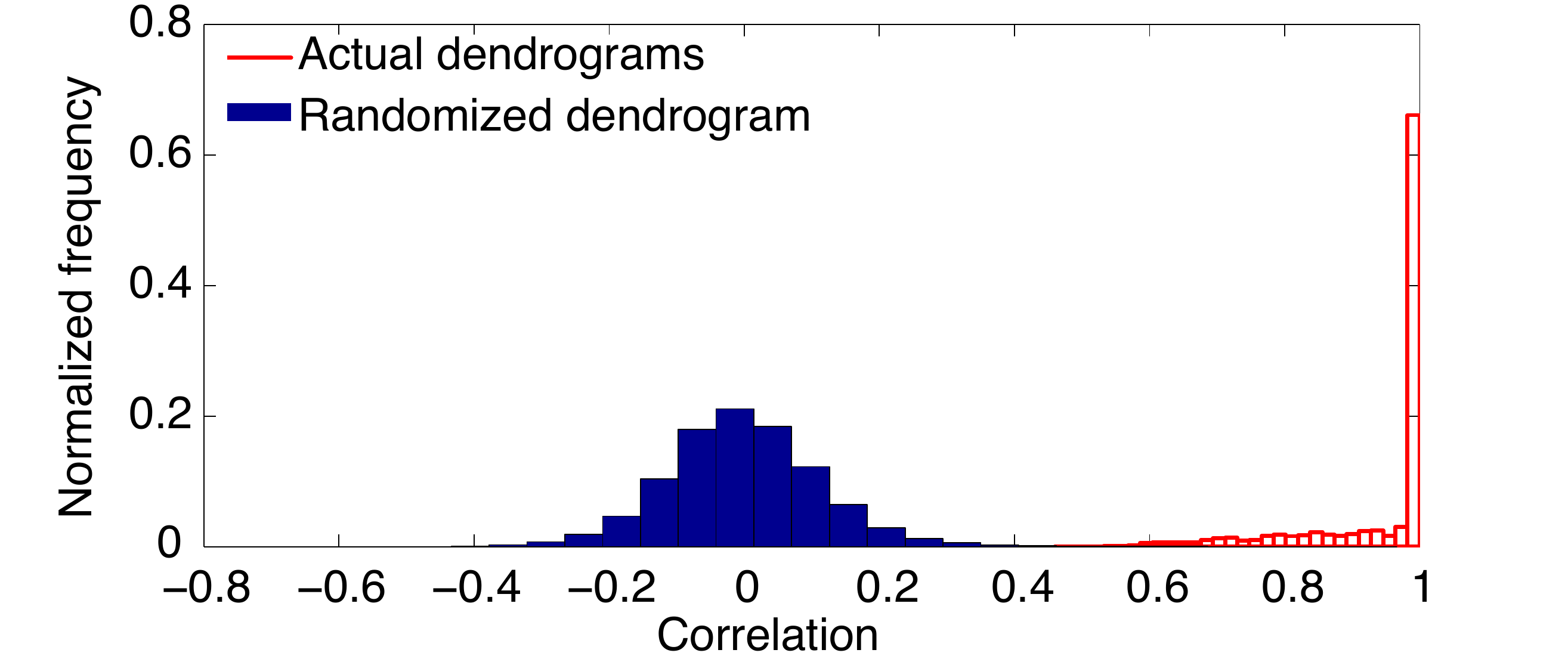}
\end{center}
\caption{(Color online)
Comparison of the distributions of correlation coefficients between empirical Louvain dendrograms and empirical (red, hollow) and randomized (blue, solid) simulated-annealing dendrograms. See the text for details.}
\label{fig:ultracorrcomp}
\end{figure}

In Fig.~\ref{fig:algo_comp_dend}, we show dendrograms obtained using the Louvain and simulated-annealing modularity optimization algorithms for a subset of 15 networks. On visual inspection, the dendrograms appear to be very similar, as there are only a few small differences in the heights at which leaves and clusters combine. To quantify the similarity between a pair of dendrograms with underlying distance matrices $\mathbf{s}$ and $\mathbf{t}$, we define a correlation coefficient $\varphi$ as
\begin{equation}
	\varphi=\frac{\sum_{i<j} \big( s_{ij}-\bar{s} \big) \big( t_{ij}-\bar{t} \big) }{\sqrt{\Big[ \sum_{i<j} \big( s_{ij}-\bar{s} \big) ^2 \Big] \Big[ \sum_{i<j} \big( t_{ij}-\bar{t} \big) ^2 \Big] }}\,,
\end{equation}
where $\bar{s}$ is the mean of the distances $s_{ij}$ and $\bar{t}$ is the mean of the distances $t_{ij}$. Dendrograms derived from identical distance matrices have correlation coefficient $\varphi=1$. The correlation for the example dendrograms shown in Fig.~\ref{fig:algo_comp_dend} is 0.997. To get a better sense of the extent of this correlation, we compare the observed correlations with those obtained for randomized dendrograms. To make the comparison, we first produce a distribution of correlation coefficients $\varphi$ between a large number of empirical (unrandomized) dendrograms produced by the Louvain and simulated-annealing algorithms. Because of the computational costs of calculating the MRFs for the simulated annealing algorithm, we only consider the subset of 25 networks identified above. We select 15 networks uniformly at random from this subset of 25 networks and generate two dendrograms similar to those in Fig.~\ref{fig:algo_comp_dend}: one corresponds to the distance matrix produced by the Louvain algorithm and the other corresponds to the distance matrix produced by simulated annealing. We then calculate the correlation coefficient between the two distance matrices. We repeat this process 10,000 times to obtain 10,000 correlation coefficients, whose distribution we show using the hollow red histogram in Fig.~\ref{fig:ultracorrcomp}. This procedure makes it possible to compare a large number of dendrograms at the computational cost of calculating simulated annealing MRFs for a total of 25 networks, highlighted with asterisks in Table 2 of the Supplemental Material.

We then compare this observed distribution of correlation coefficients to a randomized reference. We focus on the correlation between empirical Louvain dendrograms (i.e., empirical dendrograms resulting from distance matrices produced by the Louvain method) and randomized simulated-annealing dendrograms (i.e., dendrograms resulting from distance matrices produced by the simulated annealing algorithm that have been subsequently randomized). We proceed as follows: for each of the 10,000 dendrogram pairs that we assembled from subsets of 15 networks, we create 100 randomizations of the simulated-annealing dendrogram, and we then calculate the correlation coefficient between each of these randomized dendrograms and the corresponding empirical Louvain dendrogram. The resulting distribution from 10,000 repetitions is the solid blue histogram in Fig.~\ref{fig:ultracorrcomp}. To randomize the simulated-annealing dendrogram, we used the double-permutation procedure described in Refs.~\cite{DENDCOMP1,DENDCOMP2}.  This procedure has two steps. First, we randomize the distances at which the different clusters are combined. For example, consider an unrandomized dendrogram in which clusters A and B are combined at a distance of 0.45 and clusters C and D are combined at a distance of 0.65; after the randomization, A and B might be combined at a distance of 0.65 and C and D might be combined at a distance of 0.45. Second, we randomize the networks corresponding to each leaf in the dendrogram. This two-step randomization procedure maintains the underlying distances and the topology of the dendrogram.

As mentioned above, we show the distributions of correlation coefficients between empirical Louvain dendrograms and the empirical (unrandomized) and randomized simulated-annealing dendrograms in Fig.~\ref{fig:ultracorrcomp}. The correlation is clearly much higher for the empirical case, as there is only a very slight overlap in the tails of the two distributions. The correlation between the Louvain and simulated-annealing dendrograms is greater than 0.99 for about 63\% of the studied dendrograms. 

\section{Diagnostic for Assessing the Clustering from Different Distance Measures}
\label{sec:clustermetric}

An examination of the leaf colors of the dendrogram in Fig.~\ref{fig:category_MRFs} illustrates that the employed distance measure groups together networks from a variety of categories, including political voting networks, political committee networks, Facebook networks, metabolic networks, and fungal networks. A visual comparison provides a reasonable starting point for assessing the effectiveness of different distance measures at clustering networks. To quantify how effectively each distance matrix ($\mathbf{D}^\hh$, $\mathbf{D}^S$, $\mathbf{D}^{\eta}$, and $\mathbf{D}^p$) clusters networks of the same type, we introduce a \emph{clustering diagnostic}, which we denote by $\alpha(h)$, to be explained shortly. Because the assignment of networks to categories is subjective and because some of the categories include networks of very different types, it would be inappropriate to assess the effectiveness of a distance measure based on how well it clusters networks in very broad categories. We thus focus our examination on narrower categories whose constituent networks are clustered fairly tightly in Fig.~\ref{fig:category_MRFs}. This includes the following 8 categories of networks: Facebook, metabolic, political cosponsorship, political committee, political voting, financial, brain, and fungal.

The clustering diagnostic depends on where one  ``cuts" the dendrograms.  We start by constructing a dendrogram for each of the four distance matrices $\mathbf{D}^\hh$, $\mathbf{D}^S$, $\mathbf{D}^{\eta}$, and $\mathbf{D}^p$.  Performing a horizontal cut through a dendrogram at a given height $h$ splits the dendrogram into multiple disconnected clusters ($h$ is measured in terms of ultrametric distances; see Fig.~\ref{fig:algo_comp_dend}). For each such cluster, we calculate the proportion of networks from a particular category that are contained in it. For example, if a cut produces three clusters and if we consider the Facebook category, then we might find that one cluster contains two tenths of the Facebook networks, a second cluster has three tenths of those networks, and the third cluster has the remaining half of those networks. We calculate these membership fractions for each network category and for each cluster. We then identify, for each category, what we called the \emph{plurality cluster}, which is defined as the cluster that includes the largest fraction of networks from that category. In the above example, the third cluster is the plurality cluster for the Facebook category.  Our diagnostic $\alpha(h)$ is then defined by adding across all 8 categories the fraction of networks in the plurality clusters:
\begin{equation}
	\alpha(h) = \sum_{j = 1}^8 \gamma_j(h)\,,
\end{equation}
where $\gamma_j(h)$ is the plurality fraction for the $j$th category of networks for the given cut at height $h$ of the dendogram.		 

We perform similar calculations for each level of the dendrogram and use the resulting values of $\alpha(h)$ to assess the effectiveness of the different distance measures at clustering the networks. For example, at the root of the dendrogram, all of the networks are in a single cluster, so the maximum fraction of networks in the same cluster is $1$ for every network category.  Given the above choice of 8 categories, this yields $\alpha = 8$. However, as one considers lower levels of the dendrogram, the clusters break up more and more, so the fraction of networks in the plurality cluster in each category typically decreases. Effective distances measures ought to result in relatively high values for $\alpha(h)$.

\begin{figure}[tbp]
\begin{center}
\includegraphics[width=1\linewidth]{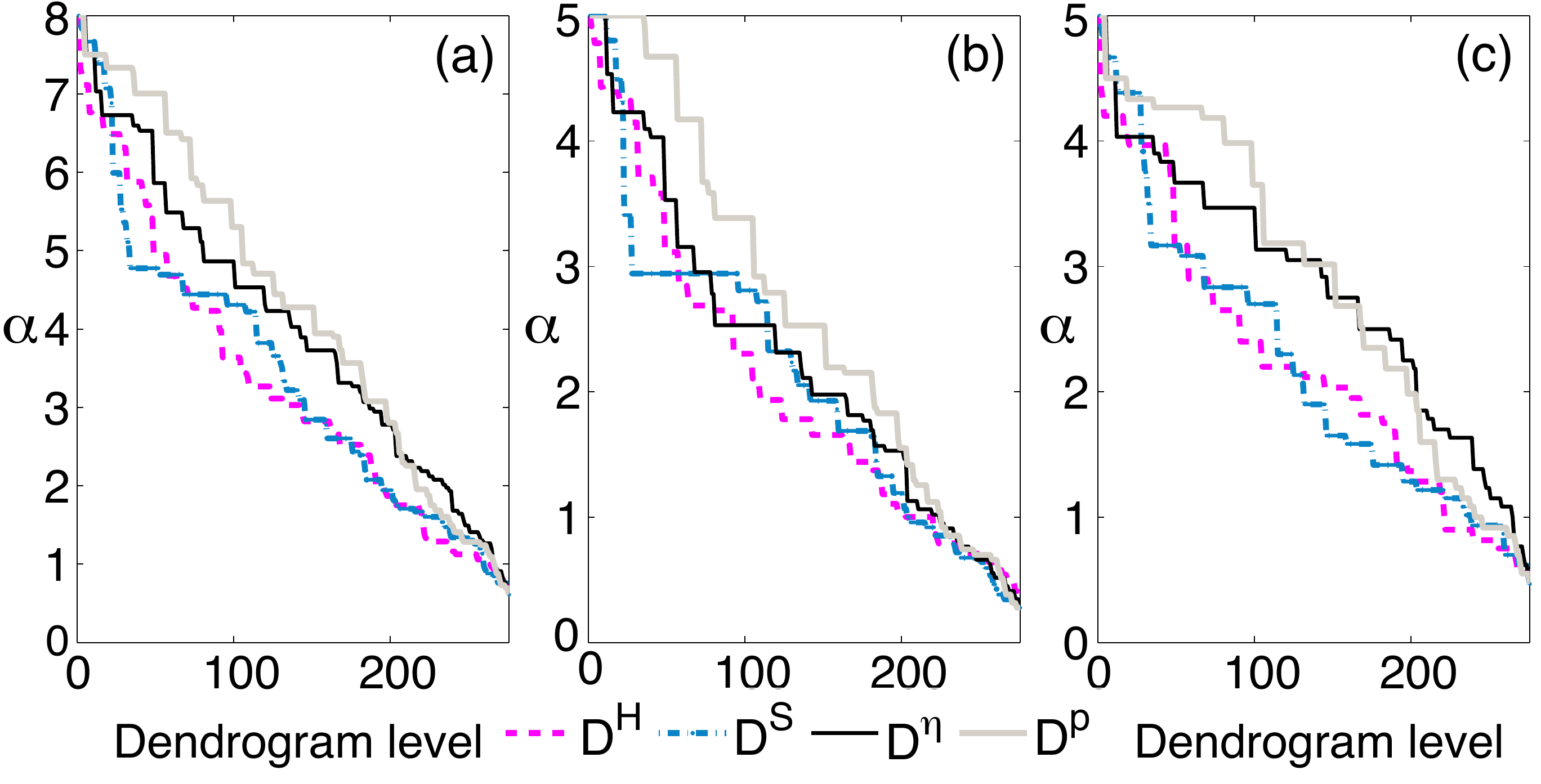}
\end{center}
\caption{(Color online)
Comparison of the effectiveness of the employed distance measures at clustering networks of the same category. As discussed in this text, we quantify this using the clustering diagnostic $\alpha(h)$.  We calculate dendrograms from four distance matrices ($\mathbf{D}^\hh$, $\mathbf{D}^S$, $\mathbf{D}^{\eta}$, and $\mathbf{D}^p$) and compare the resulting values of $\alpha(h)$ for different sets of categories. (\textbf{a}) The value of the clustering diagnostic $\alpha(h)$ as a function of dendrogram cut level $h$ (i.e., where the dendrogram is split to clusters) for the following 8 categories of networks: Facebook, metabolic, political cosponsorship, political committee, political voting, financial, brain, and fungal. (\textbf{b}) The value of $\alpha(h)$ for the largest 5 of the above 8 categories (Facebook, metabolic, political cosponsorship, political committee, and political voting) and (\textbf{c}) for the smallest 5 of the above 8 categories (Facebook, metabolic, financial, brain, and fungal). The maximum possible value of $\alpha(h)$ in each panel is equal to the number of categories considered in each panel.  The values of $\alpha(h)$ obtained using the PCA-distance matrix $\mathbf{D}^p$ (gray solid curve) are usually higher than those obtained using the other three distance measures.  This suggests that PCA distance is the most effective of the four employed clustering measures.}
\label{fig:dend_clustering_stats}
\end{figure}

In Fig.~\ref{fig:dend_clustering_stats}, we compare the values of $\alpha(h)$ at each level of the dendrogram for $\mathbf{D}^\hh$, $\mathbf{D}^S$, $\mathbf{D}^{\eta}$, and $\mathbf{D}^p$. For each of the different subsets of networks and for most of the dendrogram levels, the PCA-distance $\mathbf{D}^p$ is the most effective of the employed distance measures at clustering networks of the same category.  This agrees with our visual assessment (i.e., our identification of contiguous blocks of color) of the different measures.

\section{Using Simple Characteristics to Cluster Networks}
\label{sec:altchar}

We established in Section~\ref{sec:emptaxons} that the PCA-distances $\mathbf{D}^{p}$ between MRFs can produce sensible network taxonomies, and we now consider briefly whether the observed taxonomies can be explained using simple summary statistics.  We consider only a few specific properties, though of course there are myriad other network diagnostics that one might consider.

\begin{figure}[tbp]
\begin{center}
\includegraphics[width=1\linewidth]{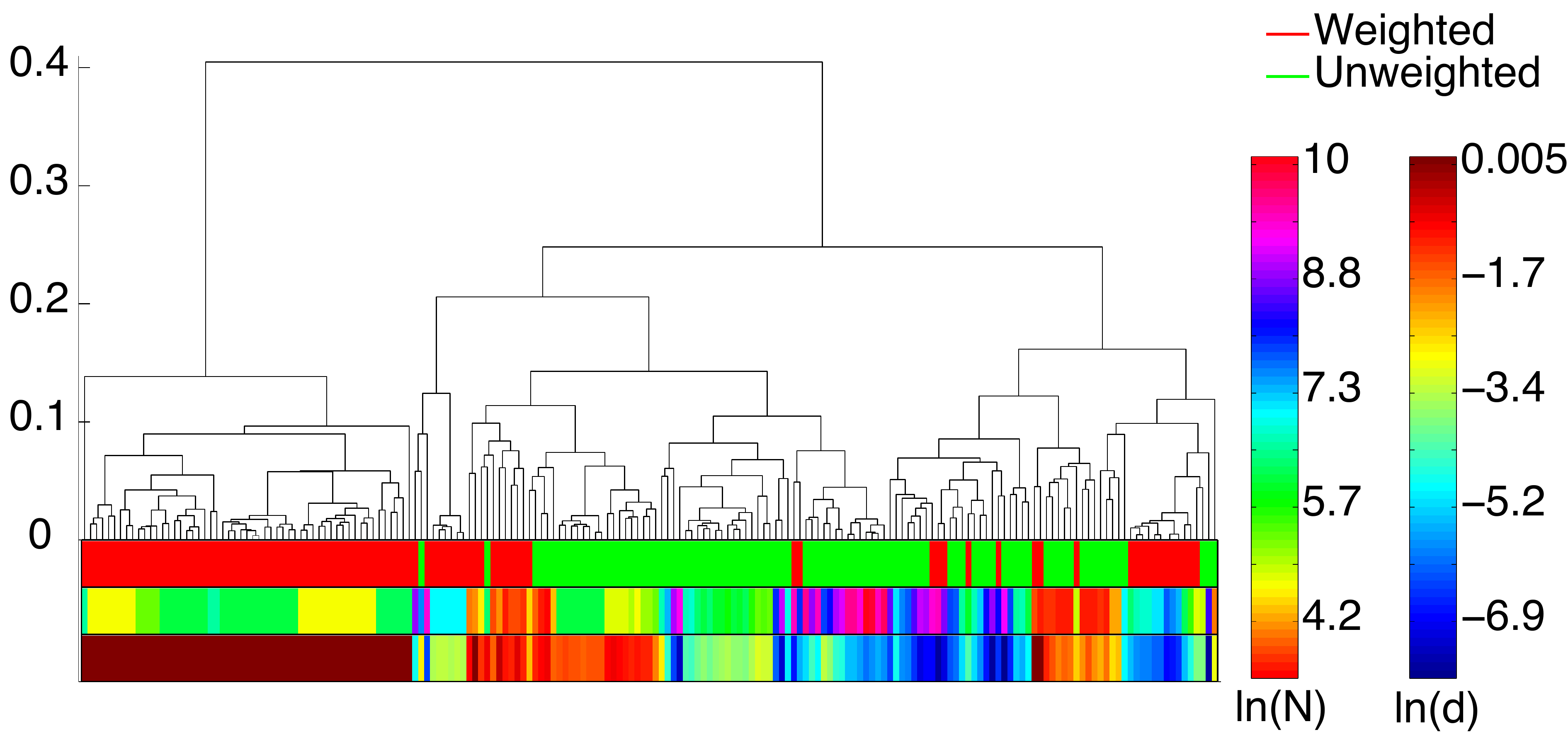}
\end{center}
\caption{(Color online)
Taxonomy for 189 networks. We constructed the dendrogram using the distance matrix $\mathbf{D}^p$ and average linkage clustering. We order the leaves of the dendrogram to minimize the distance between adjacent nodes, and we color the leaves to indicate the type of network. The three color bars below the dendrogram indicate whether the network corresponding to each leaf is weighted or unweighted (top), the number of nodes in the networks $N$ (middle), and the fraction of possible edges that are present $d$ (bottom).}
\label{fig:network_properties_dend}
\end{figure}

Perhaps the three simplest properties of an undirected network are the following: (1) whether it has weighted or unweighted edges; (2) the number of nodes $N$; and (3) the edge density $d=2L/[N(N-1)]$ (where $L$ is the number of edges, which we distintinguish from the total edge weight $m$ in weighted networks). The top colored row in Fig.~\ref{fig:network_properties_dend} indicates that many of the weighted networks are clustered together at the far left of the dendrogram. However, there are also weighted networks scattered throughout the dendrogram, so whether a network is weighted or unweighted does not explain the observed classification. The third colored row provides a clearer explanation for the cluster of networks at the left: These are not simply weighted networks, as they are in fact similarity networks, so that nearly all possible edges are present and have weights indicating connection strengths. However, this property alone cannot explain the observed classification, as several of the weighted networks containing nearly all possible edges do not appear at the far left of the dendrogram. In fact, there are many clusters in the dendrogram that contain networks with very different fractions of possible edges. The total number of nodes, shown by the second colored row in the figure, again explains some of the clustering, as networks with similar numbers of nodes are clustered together in some regions of the dendrogram. However, there are also numerous examples in which networks with the same number of nodes appear in different clusters. Therefore, none of these three simple network diagnostics can explain the observed classification by itself.

\clearpage

\setcounter{page}{1}
\renewcommand{\thepage}{SM-\arabic{page}}
\renewcommand{\figurename}{SM FIG.} 
\renewcommand{\tablename}{SM TABLE} 

\begin{center}
\section*{Supplemental Material}
\end{center}

\subsection*{Taxonomy Of All Studied Networks}
In Supplemental Fig.~\ref{fig:fulld}, we show a dendrogram containing leaves for all of the 746 networks that we studied. This dendrogram contains several large contiguous blocks of leaves that correspond to networks belonging to the same category. For example, there are large contiguous blocks of fungal, Facebook, metabolic, political committee, political voting, and financial networks. These blocks do not always include all of the networks within a category; when there are separate contiguous blocks for the same category, the blocks sometimes correspond to different types of networks within a category. For example, the political voting networks category includes separate blocks of UN voting networks and UK House of Commons voting networks. However, because of the number of networks that we include in the study and the imbalance in the spread of networks across categories, Fig.~\ref{fig:fulld} is difficult to interpret and the smaller categories are obfuscated by the larger ones. Therefore, for a clearer view of the relationships between the different categories of networks, we focus in the main text on 189 of the 746 networks.
\setcounter{figure}{0}
\begin{figure*}[!htb]
\begin{center}
\includegraphics[width=1\linewidth]{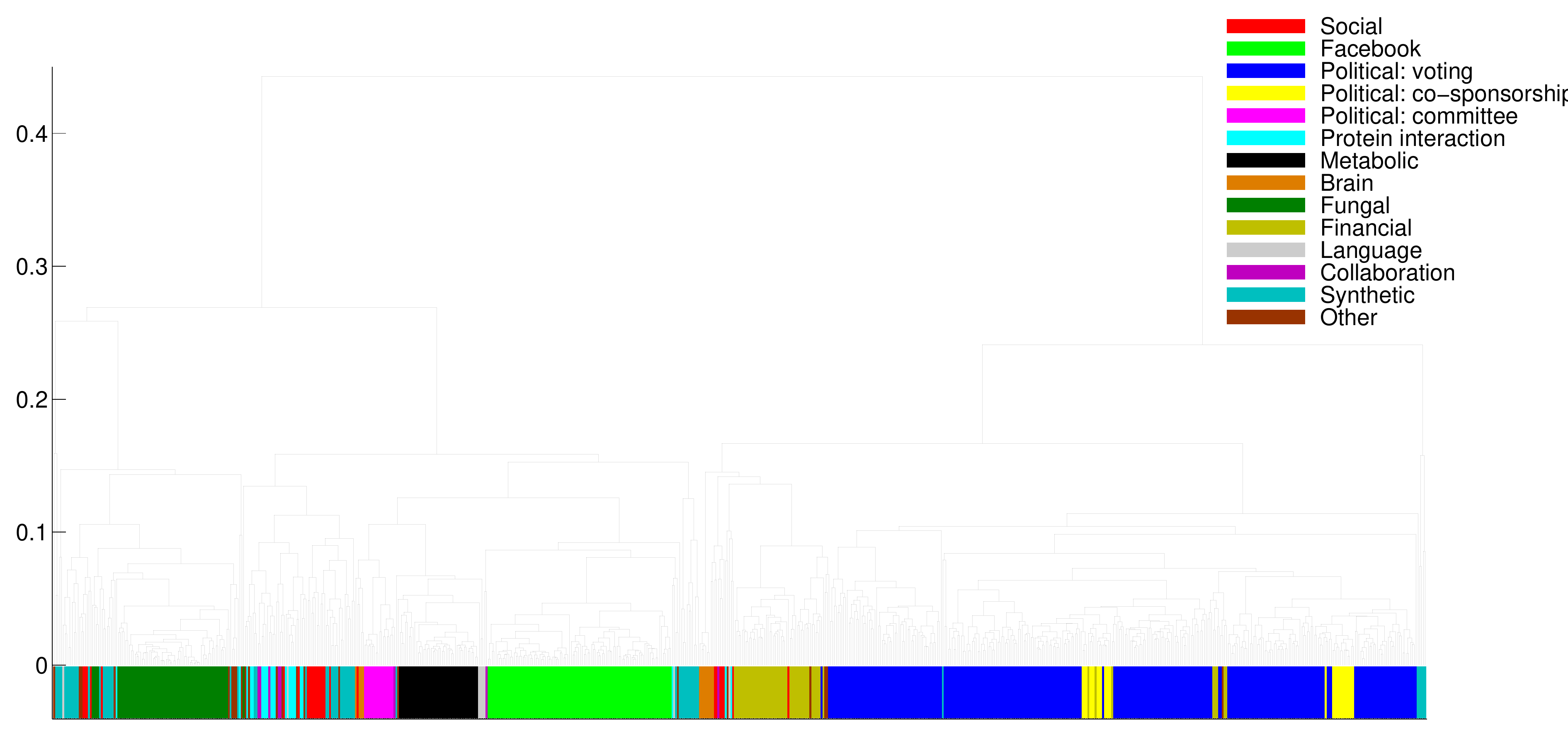}
\end{center}
\caption{(Color online) Dendrogram for 746 networks obtained using mesoscopic response functions (MRFs). Note that the colors used to indicate network categories are not the same as those in the main text.}
\label{fig:fulld}
\end{figure*}

\subsection*{Details of Networks}

In Supplemental Table II, we provide details of all of the networks that we employed in our study. The set of networks includes several several synthetic network models as well as synthetic benchmark networks that were introduced to test community detection algorithms. We include multiple realizations (using various parameter values) for many of the model and benchmark networks. In this section, we briefly describe these synthetic networks and explain the notation that we use to label them in Supplemental Table~II.

\subsection{Erd\"{o}s-R\'{e}nyi (ER)}
In an ER network of $N$ nodes, each pair of nodes is connected by an unweighted edge with probability $p$ (and is not connected with probability $1-p$) \cite{er}. The degree of each node is distributed according to a binomial distribution. We label the ER networks using the notation ``ER: ($N$,$p$)''.

\subsection{Watts-Strogatz (WS)}
We consider the small-world network of Watts and Strogatz \cite{ws} for a one-dimensional lattice of $N$ nodes with periodic boundary conditions. The network consists of a ring in which each node is connected with an unweighted edge to all of its neighbors that are $k$ or fewer lattice spacings away. Each edge is then considered in turn and one end is rewired with probability $p$ to a different node selected uniformly at random, subject to the constraint that there can be no self-edges or multi-edges. We label each Watts-Strogatz network as ``WS: ($N$,$k$,$p$)''.

\subsection{Barab\'{a}si-Albert (BA)}
BA networks \cite{ba} are obtained using a network growth mechanism in which nodes with degree $m$ are added to the network, one per time step, and the other end of each new edge attaches to an existing node with a probability proportional to the degree of that node. We label each BA network as ``BA: ($N$,$m$)''.

\subsection{Fractal}
We generate fractal networks using the method described in Ref.~\cite{fractal}. We begin by generating an isolated group of $2^\beta$ fully connected nodes, where $\beta$ gives the size of the clusters. These groups correspond to the hierarchical level $h=0$. We then create a second identical group and connect the two groups using an edge density of $f_e^{-1}$, where $f_e$ is the number of edges out of all possible edges between the groups. We then duplicate this network and connect the two duplicates at the level $h=2$ using an edge density of $f_e^{-2}$. We repeat this until we reach the desired network size $N=2^n$, where $n$ is the number of hierarchical levels. At each step, the connection density is decreased, resulting in progressively sparser interconnectivity at higher hierarchical levels. The resulting network exhibits self-similar properties. We label each network ``Fractal: ($n,\beta,f_e$)''.

\subsection{Random Fully-Connected}
We produce randomly weighted, fully connected networks of $N$ nodes by connecting every node to every other node with an edge whose weight is chosen uniformly at random on the unit interval. The networks have $N(N-1)/2$ edges. We label each network ``Random fully-connected: ($N$)''.

\subsection{Kumpula-Onnela-Saram\"{a}ki-Kaski-Kert\'{e}sz (KOSKK) model}
We generate weighted networks containing communities using the model described in Ref.~\cite{weighted}. We create edges via two mechanisms. First, at every time-step, each node $i$ selects a neighbor $j$ with probability $w_{ij}/s_i$, where $w_{ij}$ is the weight of of the edge connecting $i$ and $j$ and $s_i=\sum_j w_{ij}$ is the strength of $i$. If $j$ has other neighbors in addition to $i$, then one of them is selected with probability $w_{jk}/(s_j-w_{ij})$. If $i$ and $k$ are not connected, then a new edge of weight $w_{ik}=w_0$ is created with probability $p_n$. If the edge already exists, its weight is increased by an amount $\delta$. In both cases, $w_{ij}$ and $w_{jk}$ are also increased by $\delta$. This process is termed \textit{local attachment}. Second, if a node has no edges, then with probability $p_r$ it creates an edge of weight $w_0$ to a randomly selected node.  (This is called \emph{global attachment}.) A node can be deleted with probability $p_d$; if this happens, then all of its edges are also removed and the node is replaced by a new node, so that the total number of nodes $N$ and the mean number of edges both remain constant. We label each network ``Weighted: ($N,w_0,\delta,p_n,p_r,p_d,t$)'', where $t$ is the total number of simulation time steps.

\subsection{Lancichinetti-Fortunato-Radicchi (LFR) benchmark}
The LFR benchmark \cite{fortunato_bench} consists of unweighted networks with non-overlapping communities. A network in this ensemble is constructed by assigning each node a degree from a power-law distribution with exponent $\gamma$, where the extremes of the distribution $k_{\min}$ and $k_{\max}$ are chosen so that the mean degree is $\langle k \rangle$, and the nodes are connected using the configuration model \cite{configmodel}. Each node shares a fraction $\mu$ of its edges with nodes in other communities and a fraction $1-\mu$ of them with nodes in its own community. The community sizes are taken from a power-law distribution with exponent $\beta$, subject to the constraint that the sum of all of the community sizes equals the number of nodes $N$ in the network. The minimum and maximum community sizes ($q_{\min}$ and $q_{\max}$) are then chosen to satisfy the additional constraint that $q_{\min}>k_{\min}$ and $q_{\max}>k_{\max}$, which ensures that each node is included in at least one community. We label each network ``LFR: ($N,\langle k \rangle,k_{\max},\gamma,\beta,\mu,q_{\min},q_{\max}$)''.

\subsection{Lancichinetti-Fortunato (LF) benchmark}
The LF benchmark \cite{fortunato_bench2} allows networks to be weighted and the communities to overlap. In the present paper, we only consider weighted networks with non-overlapping communities. The node degrees are again taken from a power-law degree distribution (as in LFR benchmark networks), but this time we label the exponent $\tau_1$, and the community sizes are taken from a power-law degree distribution with exponent $\tau_2$. The strength $s_i$ of each node is chosen so that $s_i=k_i^\beta$, where $k_i$ again gives the degree of node $i$. There are also two mixing parameters: a topological mixing parameter $\mu_t$, which specifies the proportion of edges outside a node's community; and a participation mixing parameter $\mu_w$, which specifies the weight of a node's edges outside its community. We label each network ``LF: ($N,\langle k \rangle,k_{\max},\mu_t,\mu_w,\beta,\tau_1,\tau_2$)''. For all of the LF networks, we set $N=1000$. One can alternatively set the minimum and maximum community sizes $q_{\min}$ and $q_{\max}$. We always use $q_{\min}=20$ and $q_{\max}=50$, so we do not include these parameters when we label the networks.

\subsection{LF-Newman-Girvan benchmark}
We include an LF network ensemble with parameters values $N=128$, $\langle k \rangle=16$, $k_{\max}=16$, $\mu_w=0.1$, $q_{\min}=32$, $q_{\max}=32$, and $\beta=1$.  This family of networks is similar to the NG benchmark \cite{newmangirvan,fortunato_bench2}.

\setcounter{table}{0}
\begin{table}[!htp]
\caption{\label{tbl:randsubsets}The networks included in the 6 subsets of 25 networks used to test the robustness of the clusterings to random perturbations in Appendix A of the main text. The network ID corresponds to the numerical identifier of the network in Supplemental Table~II.}
\begin{center}
\begin{tabular}{|c|p{15cm}|}
\hline
Subset	& Network IDs\\
\hline
\hline
1 & \scriptsize{9, 11, 243, 251, 267, 269, 280, 283, 301, 305, 340, 351, 353, 646, 662, 665, 674, 688, 690, 693, 700, 711, 735, 736, 740} \\
2 & \scriptsize{71, 250, 251, 252, 264, 267, 270, 271, 276, 285, 301, 303, 305, 347, 354, 646, 649, 669, 674, 683, 688, 691, 737, 738, 739} \\
3 & \scriptsize{11, 12, 243, 254, 264, 266, 268, 273, 284, 290, 291, 302, 308, 340, 341, 354, 355, 645, 649, 662, 672, 690, 695, 697, 717} \\
4 & \scriptsize{252, 253, 258, 261, 264, 268, 274, 279, 280, 286, 288, 291, 342, 347, 348, 352, 641, 669, 694, 696, 699, 700, 711, 714, 737} \\
5 & \scriptsize{10, 20, 34, 248, 250, 261, 266, 268, 272, 283, 285, 301, 306, 308, 344, 656, 661, 666, 693, 694, 700, 709, 711, 712, 715} \\
6 & \scriptsize{34, 249, 252, 253, 256, 257, 266, 277, 283, 289, 291, 301, 302, 340, 341, 345, 650, 656, 661, 662, 690, 709, 713, 717, 736} \\
\hline
\end{tabular}
\end{center}
\end{table}

\setcounter{table}{1}
\begin{table*}
\caption[Network details and statistics]{Network summary statistics. We symmetrize all networks, remove self-edges, and only consider largest connected components. In this table, we give the network category, whether it is weighted or unweighted, the number of nodes $N$ in the largest connected component, the number of edges $L$ in this component, the fraction of possible edges present $f_e=2L/[N(N-1)]$, and a reference providing details of the data source. We highlight the 25 networks used in the randomizations in Appendix A in bold and the 189 networks used in the aggregate taxonomy in red. We indicate with an asterisk ($\ast$) all networks used in Appendix B.2 to test the robustness of the taxonomy to different optimization heuristics.}
\begin{ruledtabular}
\begin{tabular}{|c|l|l|c|c|c|c|c|}
ID & Name & Category & Weighted & $N$ & $L$ & $f_e$ & References \\ 
\hline
1 & \textcolor{red}{Human brain cortex: participant A1} & Brain & Y & 994 & 13,520 & 0.0274 & \cite{human_brain} \\
2 & \textcolor{red}{Human brain cortex: participant A2} & Brain & Y & 987 & 14,865 & 0.0305 & \cite{human_brain} \\
3 & \textcolor{red}{Human brain cortex: participant B} & Brain & Y & 980 & 14,222 & 0.0296 & \cite{human_brain} \\
4 & \textcolor{red}{Human brain cortex: participant D} & Brain & Y & 996 & 14,851 & 0.0300 & \cite{human_brain} \\
5 & \textcolor{red}{Human brain cortex: participant E} & Brain & Y & 992 & 14,372 & 0.0292 & \cite{human_brain} \\
6 & \textcolor{red}{Human brain cortex: participant C} & Brain & Y & 996 & 14,933 & 0.0301 & \cite{human_brain} \\
7 & \textcolor{red}{Cat brain: cortical}$^*$ & Brain & Y & 52 & 515 & 0.3884 & \cite{cat_brain} \\
\end{tabular}
\end{ruledtabular}
\end{table*}

\setcounter{table}{1}
\begin{table*}
\caption{(Continued.)}
\begin{ruledtabular}
\begin{tabular}{|c|l|l|c|c|c|c|c|}
ID & Name & Category & Weighted & $N$ & $L$ & $f_e$ & References \\
\hline
51 & Facebook: Johns & Facebook & N & 5,157 & 186,572 & 0.0140 & \cite{traud} \\
52 & Facebook: Lehigh & Facebook & N & 5,073 & 198,346 & 0.0154 & \cite{traud} \\
53 & Facebook: Maine & Facebook & N & 9,065 & 243,245 & 0.0059 & \cite{traud} \\
54 & Facebook: Maryland & Facebook & N & 20,829 & 744,832 & 0.0034 & \cite{traud} \\
55 & Facebook: Mich & Facebook & N & 3,745 & 81,901 & 0.0117 & \cite{traud} \\
56 & Facebook: Michigan & Facebook & N & 30,106 & 1,176,489 & 0.0026 & \cite{traud} \\
57 & Facebook: Middlebury & Facebook & N & 3,069 & 124,607 & 0.0265 & \cite{traud} \\
\end{tabular}
\end{ruledtabular}
\end{table*}

\setcounter{table}{1}
\begin{table*}
\caption{(Continued.)}
\begin{ruledtabular}
\begin{tabular}{|c|l|l|c|c|c|c|c|}
ID & Name & Category & Weighted & $N$ & $L$ & $f_e$ & References \\
\hline
101 & Facebook: UF & Facebook & N & 35,111 & 1,465,654 & 0.0024 & \cite{traud} \\
102 & Facebook: UGA & Facebook & N & 24,380 & 1,174,051 & 0.0040 & \cite{traud} \\
103 & \textcolor{red}{Facebook: UNC} & Facebook & N & 18,158 & 766,796 & 0.0047 & \cite{traud} \\
104 & Facebook: USC & Facebook & N & 17,440 & 801,851 & 0.0053 & \cite{traud} \\
105 & \textcolor{red}{Facebook: USF} & Facebook & N & 13,367 & 321,209 & 0.0036 & \cite{traud} \\
106 & Facebook: USFCA & Facebook & N & 2,672 & 65,244 & 0.0183 & \cite{traud} \\
107 & Facebook: UVA & Facebook & N & 17,178 & 789,308 & 0.0054 & \cite{traud} \\
\end{tabular}
\end{ruledtabular}
\end{table*}

\setcounter{table}{1}
\begin{table*}
\caption{(Continued.)}
\begin{ruledtabular}
\begin{tabular}{|c|l|l|c|c|c|c|c|}
ID & Name & Category & Weighted & $N$ & $L$ & $f_e$ & References \\
\hline
151 & NYSE: H1 1997 & Financial & Y & 100 & 4,950 & 1.0000 & \cite{nysedjf} \\
152 & NYSE: H2 1997 & Financial & Y & 100 & 4,950 & 1.0000 & \cite{nysedjf} \\
153 & NYSE: H1 1998 & Financial & Y & 100 & 4,950 & 1.0000 & \cite{nysedjf} \\
154 & NYSE: H2 1998 & Financial & Y & 100 & 4,950 & 1.0000 & \cite{nysedjf} \\
155 & NYSE: H1 1999 & Financial & Y & 100 & 4,950 & 1.0000 & \cite{nysedjf} \\
156 & NYSE: H2 1999 & Financial & Y & 100 & 4,950 & 1.0000 & \cite{nysedjf} \\
157 & NYSE: H1 2000 & Financial & Y & 100 & 4,950 & 1.0000 & \cite{nysedjf} \\
\end{tabular}
\end{ruledtabular}
\end{table*}

\setcounter{table}{1}
\begin{table*}
\caption{(Continued.)}
\begin{ruledtabular}
\begin{tabular}{|c|l|l|c|c|c|c|c|}
ID & Name & Category & Weighted & $N$ & $L$ & $f_e$ & References \\
\hline
201 & \textcolor{red}{\emph{Phanerochaete velutina} control11-11}$^*$ & Fungal & Y & 823 & 954 & 0.0028 & \cite{heaton} \\
202 & \emph{Phanerochaete velutina} control17-1 & Fungal & Y & 16 & 15 & 0.1250 & \cite{heaton} \\
203 & \textcolor{red}{\emph{Phanerochaete velutina} control17-2} & Fungal & Y & 232 & 240 & 0.0090 & \cite{heaton} \\
204 & \emph{Phanerochaete velutina} control17-3 & Fungal & Y & 502 & 539 & 0.0043 & \cite{heaton} \\
205 & \emph{Phanerochaete velutina} control17-4 & Fungal & Y & 703 & 754 & 0.0031 & \cite{heaton} \\
206 & \textcolor{red}{\emph{Phanerochaete velutina} control17-5} & Fungal & Y & 816 & 874 & 0.0026 & \cite{heaton} \\
207 & \emph{Phanerochaete velutina} control17-6 & Fungal & Y & 950 & 1,058 & 0.0023 & \cite{heaton} \\
\end{tabular}
\end{ruledtabular}
\end{table*}

\setcounter{table}{1}
\begin{table*}
\caption{(Continued.)}
\begin{ruledtabular}
\begin{tabular}{|c|l|l|c|c|c|c|c|}
ID & Name & Category & Weighted & $N$ & $L$ & $f_e$ & References \\
\hline
251 & \textbf{Metabolic: AP} & Metabolic & N & 201 & 857 & 0.0426 & \cite{metabolic} \\
252 & Metabolic: AT & Metabolic & N & 296 & 1,231 & 0.0282 & \cite{metabolic} \\
253 & Metabolic: BB & Metabolic & N & 175 & 628 & 0.0412 & \cite{metabolic} \\
254 & Metabolic: BS & Metabolic & N & 772 & 3,611 & 0.0121 & \cite{metabolic} \\
255 & Metabolic: CA & Metabolic & N & 483 & 2,274 & 0.0195 & \cite{metabolic} \\
256 & \textcolor{red}{Metabolic: CE} & Metabolic & N & 453 & 2,025 & 0.0198 & \cite{metabolic} \\
257 & Metabolic: CJ & Metabolic & N & 370 & 1,631 & 0.0239 & \cite{metabolic} \\
\end{tabular}
\end{ruledtabular}
\end{table*}

\setcounter{table}{1}
\begin{table*}
\caption{(Continued.)}
\begin{ruledtabular}
\begin{tabular}{|c|l|l|c|c|c|c|c|}
ID & Name & Category & Weighted & $N$ & $L$ & $f_e$ & References \\
\hline
301 & \textbf{Garfield: small-world citations} & Other & N & 233 & 994 & 0.0368 & \cite{garfield} \\
302 & Electronic circuit (s208)$^*$ & Other & N & 122 & 189 & 0.0256 & \cite{superfamily} \\
303 & Electronic circuit (s420) & Other & N & 252 & 399 & 0.0126 & \cite{superfamily} \\
304 & Electronic circuit (s838) & Other & N & 512 & 819 & 0.0063 & \cite{superfamily} \\
305 & \textbf{Protein: serine protease inhibitor (1EAW)}$^*$ & Other & N & 53 & 123 & 0.0893 & \cite{superfamily} \\
306 & Protein: immunoglobulin (1A4J)$^*$ & Other & N & 95 & 213 & 0.0477 & \cite{superfamily} \\
307 & Protein: oxidoreductase (1AOR)$^*$ & Other & N & 97 & 212 & 0.0455 & \cite{superfamily} \\
\end{tabular}
\end{ruledtabular}
\end{table*}

\setcounter{table}{1}
\begin{table*}
\caption{(Continued.)}
\begin{ruledtabular}
\begin{tabular}{|c|l|l|c|c|c|c|c|}
ID & Name & Category & Weighted & $N$ & $L$ & $f_e$ & References \\
\hline
351 & \textcolor{red}{\textbf{Committees: US House 104, Reps.}} & Political: committee & N & 432 & 17,130 & 0.1840 & \cite{map_cong,map_cong2} \\
352 & \textcolor{red}{Committees: US House 105, Reps.} & Political: committee & N & 435 & 18,297 & 0.1938 & \cite{map_cong,map_cong2} \\
353 & \textcolor{red}{\textbf{Committees: US House 106, Reps.}} & Political: committee & N & 435 & 18,832 & 0.1995 & \cite{map_cong,map_cong2} \\
354 & \textcolor{red}{Committees: US House 107, Reps.} & Political: committee & N & 434 & 19,824 & 0.2110 & \cite{map_cong,map_cong2} \\
355 & \textcolor{red}{Committees: US House 108, Reps.} & Political: committee & N & 437 & 21,214 & 0.2227 & \cite{map_cong,map_cong2} \\
356 & Roll call: US House 1 & Political: voting & Y & 66 & 2,122 & 0.9893 & \cite{waugh,poole,mccarty} \\
357 & Roll call: US House 2 & Political: voting & Y & 71 & 2,428 & 0.9771 & \cite{waugh,poole,mccarty} \\
\end{tabular}
\end{ruledtabular}
\end{table*}

\setcounter{table}{1}
\begin{table*}
\caption{(Continued.)}
\begin{ruledtabular}
\begin{tabular}{|c|l|l|c|c|c|c|c|}
ID & Name & Category & Weighted & $N$ & $L$ & $f_e$ & References \\
\hline
401 & Roll call: US House 46 & Political: voting & Y & 301 & 44,987 & 0.9964 & \cite{waugh,poole,mccarty} \\
402 & Roll call: US House 47 & Political: voting & Y & 306 & 46,214 & 0.9903 & \cite{waugh,poole,mccarty} \\
403 & Roll call: US House 48 & Political: voting & Y & 338 & 56,484 & 0.9918 & \cite{waugh,poole,mccarty} \\
404 & Roll call: US House 49 & Political: voting & Y & 330 & 54,160 & 0.9977 & \cite{waugh,poole,mccarty} \\
405 & Roll call: US House 50 & Political: voting & Y & 326 & 52,907 & 0.9987 & \cite{waugh,poole,mccarty} \\
406 & Roll call: US House 51 & Political: voting & Y & 347 & 59,303 & 0.9879 & \cite{waugh,poole,mccarty} \\
407 & Roll call: US House 52 & Political: voting & Y & 340 & 57,285 & 0.9940 & \cite{waugh,poole,mccarty} \\
\end{tabular}
\end{ruledtabular}
\end{table*}

\setcounter{table}{1}
\begin{table*}
\caption{(Continued.)}
\begin{ruledtabular}
\begin{tabular}{|c|l|l|c|c|c|c|c|}
ID & Name & Category & Weighted & $N$ & $L$ & $f_e$ & References \\
\hline
451 & Roll call: US House 96 & Political: voting & Y & 440 & 96,379 & 0.9979 & \cite{waugh,poole,mccarty} \\
452 & Roll call: US House 97 & Political: voting & Y & 442 & 96,761 & 0.9928 & \cite{waugh,poole,mccarty} \\
453 & Roll call: US House 98 & Political: voting & Y & 439 & 95,922 & 0.9977 & \cite{waugh,poole,mccarty} \\
454 & Roll call: US House 99 & Political: voting & Y & 439 & 95,875 & 0.9972 & \cite{waugh,poole,mccarty} \\
455 & Roll call: US House 100 & Political: voting & Y & 440 & 96,544 & 0.9996 & \cite{waugh,poole,mccarty} \\
456 & \textcolor{red}{Roll call: US House 101} & Political: voting & Y & 440 & 96,505 & 0.9992 & \cite{waugh,poole,mccarty} \\
457 & \textcolor{red}{Roll call: US House 102} & Political: voting & Y & 441 & 96,811 & 0.9978 & \cite{waugh,poole,mccarty} \\
\end{tabular}
\end{ruledtabular}
\end{table*}

\setcounter{table}{1}
\begin{table*}
\caption{(Continued.)}
\begin{ruledtabular}
\begin{tabular}{|c|l|l|c|c|c|c|c|}
ID & Name & Category & Weighted & $N$ & $L$ & $f_e$ & References \\
\hline
501 & Roll call: US Senate 36 & Political: voting & Y & 70 & 2,370 & 0.9814 & \cite{waugh,poole,mccarty} \\
502 & Roll call: US Senate 37 & Political: voting & Y & 70 & 2,051 & 0.8493 & \cite{waugh,poole,mccarty} \\
503 & Roll call: US Senate 38 & Political: voting & Y & 54 & 1,402 & 0.9797 & \cite{waugh,poole,mccarty} \\
504 & Roll call: US Senate 39 & Political: voting & Y & 59 & 1,610 & 0.9410 & \cite{waugh,poole,mccarty} \\
505 & Roll call: US Senate 40 & Political: voting & Y & 69 & 2,274 & 0.9693 & \cite{waugh,poole,mccarty} \\
506 & Roll call: US Senate 41 & Political: voting & Y & 80 & 3,084 & 0.9759 & \cite{waugh,poole,mccarty} \\
507 & Roll call: US Senate 42 & Political: voting & Y & 75 & 2,773 & 0.9993 & \cite{waugh,poole,mccarty} \\
\end{tabular}
\end{ruledtabular}
\end{table*}

\setcounter{table}{1}
\begin{table*}
\caption{(Continued.)}
\begin{ruledtabular}
\begin{tabular}{|c|l|l|c|c|c|c|c|}
ID & Name & Category & Weighted & $N$ & $L$ & $f_e$ & References \\
\hline
551 & Roll call: US Senate 86 & Political: voting & Y & 103 & 5,246 & 0.9987 & \cite{waugh,poole,mccarty} \\
552 & Roll call: US Senate 87 & Political: voting & Y & 105 & 5,444 & 0.9971 & \cite{waugh,poole,mccarty} \\
553 & Roll call: US Senate 88 & Political: voting & Y & 103 & 5,249 & 0.9992 & \cite{waugh,poole,mccarty} \\
554 & Roll call: US Senate 89 & Political: voting & Y & 103 & 5,247 & 0.9989 & \cite{waugh,poole,mccarty} \\
555 & Roll call: US Senate 90 & Political: voting & Y & 101 & 5,048 & 0.9996 & \cite{waugh,poole,mccarty} \\
556 & Roll call: US Senate 91 & Political: voting & Y & 102 & 5,148 & 0.9994 & \cite{waugh,poole,mccarty} \\
557 & Roll call: US Senate 92 & Political: voting & Y & 102 & 5,147 & 0.9992 & \cite{waugh,poole,mccarty} \\
\end{tabular}
\end{ruledtabular}
\end{table*}

\setcounter{table}{1}
\begin{table*}
\caption{(Continued.)}
\begin{ruledtabular}
\begin{tabular}{|c|l|l|c|c|c|c|c|}
ID & Name & Category & Weighted & $N$ & $L$ & $f_e$ & References \\
\hline
601 & UN resolutions 24 & Political: voting & Y & 126 & 7,851 & 0.9970 & \cite{voeten} \\
602 & UN resolutions 25 & Political: voting & Y & 126 & 7,868 & 0.9991 & \cite{voeten} \\
603 & UN resolutions 26 & Political: voting & Y & 132 & 8,641 & 0.9994 & \cite{voeten} \\
604 & UN resolutions 27 & Political: voting & Y & 132 & 8,646 & 1.0000 & \cite{voeten} \\
605 & UN resolutions 28 & Political: voting & Y & 134 & 8,905 & 0.9993 & \cite{voeten} \\
606 & UN resolutions 29 & Political: voting & Y & 137 & 9,202 & 0.9878 & \cite{voeten} \\
607 & UN resolutions 30 & Political: voting & Y & 143 & 10,117 & 0.9965 & \cite{voeten} \\
\end{tabular}
\end{ruledtabular}
\end{table*}

\setcounter{table}{1}
\begin{table*}
\caption{(Continued.)}
\begin{ruledtabular}
\begin{tabular}{|c|l|l|c|c|c|c|c|}
ID & Name & Category & Weighted & $N$ & $L$ & $f_e$ & References \\
\hline
651 & \textcolor{red}{DIP: \textit{M. musculus}} & Protein interaction & N & 50 & 55 & 0.0449 & \cite{dip1,dip2} \\
651 & \textcolor{red}{DIP: \textit{C. elegans}} & Protein interaction & N & 2,386 & 3,825 & 0.0013 & \cite{dip1,dip2} \\
653 & \textcolor{red}{Human: Ccsb} & Protein interaction & N & 1,307 & 2,483 & 0.0029 & \cite{ccsb} \\
654 & \textcolor{red}{Human: Ophid} & Protein interaction & N & 5,464 & 23,238 & 0.0016 & \cite{ophid1,ophid2} \\
655 & \textcolor{red}{STRING: \textit{C. elegans}} & Protein interaction & N & 1,762 & 95,227 & 0.0614 & \cite{string} \\
656 & \textcolor{red}{STRING: \textit{S. cerevisiae}} & Protein interaction & N & 534 & 57,672 & 0.4053 & \cite{string} \\
657 & \textcolor{red}{Yeast: Oxford Statistics} & Protein interaction & N & 2,224 & 6,609 & 0.0027 & \cite{sumeet} \\
\end{tabular}
\end{ruledtabular}
\end{table*}

\setcounter{table}{1}
\begin{table*}
\caption{(Continued.)}
\begin{ruledtabular}
\begin{tabular}{|c|l|l|c|c|c|c|c|}
ID & Name & Category & Weighted & $N$ & $L$ & $f_e$ & References \\
\hline
701 & Fractal: (10,2,1) & Synthetic & N & 1,024 & 9,256 & 0.0177 & \cite{fractal} \\
702 & Fractal: (10,2,2) & Synthetic & N & 1,024 & 16,875 & 0.0322 & \cite{fractal} \\
703 & Fractal: (10,2,3) & Synthetic & N & 1,024 & 30,344 & 0.0579 & \cite{fractal} \\
704 & Fractal: (10,2,4) & Synthetic & N & 1,024 & 53,009 & 0.1012 & \cite{fractal} \\
705 & Fractal: (10,2,5) & Synthetic & N & 1,024 & 89,812 & 0.1715 & \cite{fractal} \\
706 & Fractal: (10,2,6) & Synthetic & N & 1,024 & 147,784 & 0.2822 & \cite{fractal} \\
707 & Fractal: (10,2,7) & Synthetic & N & 1,024 & 232,794 & 0.4445 & \cite{fractal} \\
\end{tabular}
\end{ruledtabular}
\end{table*}

\end{document}